\documentclass[showpacs,superscriptaddress,twocolumn]{revtex4-1}
\usepackage[applemac]{inputenc}
\usepackage[english]{babel}
\usepackage{amsmath,amssymb}
\usepackage[pdftex]{graphicx}
\usepackage{color}
\usepackage{array}
\usepackage{ulem}

\begin{document}

\title{Classical chaos in atom-field systems}
\author{J. Ch\'avez-Carlos}
\affiliation{Instituto de Ciencias Nucleares, Universidad Nacional Aut\'onoma de M\'exico,
Apdo. Postal 70-543, Cd. Mx., M\'exico, C.P. 04510}
\author{M. A. Bastarrachea-Magnani}
\affiliation{Instituto de Ciencias Nucleares, Universidad Nacional Aut\'onoma de M\'exico,
Apdo. Postal 70-543, Cd. Mx., M\'exico, C.P. 04510}
\author{S. Lerma-Hern\'andez}
\affiliation{Facultad  de F\'\i sica, Universidad Veracruzana,
Circuito Aguirre Beltr\'an s/n, Xalapa, Veracruz, M\'exico, C.P. 91000}
\author{J. G. Hirsch}
\affiliation{Instituto de Ciencias Nucleares, Universidad Nacional Aut\'onoma de M\'exico,
Apdo. Postal 70-543,  Cd. Mx., M\'exico, C.P. 04510}

\begin{abstract}
The relation between the onset of chaos and critical phenomena, like Quantum Phase Transitions (QPT) and Excited-State Quantum Phase transitions (ESQPT), is analyzed for atom-field systems. While it has been speculated that the onset of hard chaos is associated with ESQPT based in the resonant case, the off-resonant cases,
and a close look at the vicinity of the QPT in resonance,
 show clearly that both phenomena, ESQPT and chaos, respond to different mechanisms. The results are supported in a detailed numerical study of the dynamics of the semiclassical Hamiltonian of the Dicke model. The appearance of chaos is quantified calculating the largest Lyapunov exponent for a wide sample of initial conditions in the whole available phase space for a given energy. The percentage of the available phase space with chaotic trajectories is evaluated as a function of energy and coupling between the qubit and bosonic part, allowing to obtain maps in the space of coupling and energy, where ergodic properties are observed in the model. Different sets of Hamiltonian parameters are considered, including resonant and off-resonant cases.
\end{abstract}
\maketitle

\section{Introduction}
The Dicke model, proposed long ago \cite{Dicke54}, describes the interaction between a set of identical two-level systems (qubits) and a bosonic mode. Originally, it was proposed to describe schematically the interaction between matter and radiation, however, nowadays it has found a richer range of applicability to describe many systems from QED circuits up to Bose Einstein condensates \cite{Sche03,Sche07,Blais04,Fink09,Bau10}. 

The model exhibits interesting quantum critical phenomena: a Quantum Phase Transition(QPT), from a normal to a superradiant phase, when the atom-photon coupling reaches a critical value, and Excited-State Quantum Phase Transitions (ESQPT), at well defined values of the excitation energy. A semiclassical analysis allows  %
to understand and classify without ambiguities these phase transitions.
The QPT is a consequence of a change in the minimal energy configuration. It can be detected as a discontinuity in the second derivative of the minimal energy as a function of the qubit-boson coupling \cite{Nah13}. The ESQPT, on the other hand, is associated with a drastic change in the available phase space volume, and can be detected by a logarithmic divergence (in the superradiant phase) or a discontinuity (in both phases) in the first derivative of the density of states (DoS) as a function of energy \cite{Str14,Str15}. 

The classical Dicke Hamiltonian posses only two degrees of freedom, being the energy the sole constant of motion. Consequently, it corresponds to a non-integrable system and it would be expected the presence of chaos in the model. These features, non-integrability and quantum chaos, are inherited by the quantum realm.
Previous studies, based on the statistical analysis of the fluctuations in the quantum spectrum, suggest that the onset of chaos is related with the QPT \cite{Emary03}, or rather with the ESQPT \cite{Per11}. 
Recently, some of us were able to study noticeably larger systems using an efficient basis, casting doubts on the correlation between the onset of chaos and the critical phenomena \cite{Bas14A,Bas14B,Bas15}. In these works it was shown that the semiclassical DoS closely describes the central trend of the quantum DoS, allowing for a parameter free unfolding of the energy spectrum. The nearest neighbor energy distributions follow the Wigner surmise (typical of quantum chaotic systems) in the same energy regions where the classical dynamics is fully chaotic. This correspondence was extended to individual quantum states in Ref.\cite{Bas16}, where it was shown that the Participation Ratio of a given quantum coherent state, associated with a point in the classical phase space, spanned in the Hamiltonian eigenstate basis, scales differently with the number of qubits, depending on whether the classical trajectories are regular or chaotic. The excellent agreement between the classical and quantum models exposes a huge richness in terms of chaos and regularity. Therefore, as it would help as a guideline of the onset of chaos in the quantum model, it is worth to explore and quantify chaos in the correspondent classical Hamiltonian.

In the present contribution a detailed study of the presence of chaos in the semiclassical version of the model is presented. The analysis is performed by calculating the Lyapunov exponents along the whole available phase space. From this sampling, the percentage of chaos in the available phase space is estimated for a wide range of excitation energies and Hamiltonian parameters, unveiling all the regimes present in the model. The analysis includes the resonant and two off-resonance cases. Complete charts of chaoticity are obtained for these different sets, allowing to identify the ergodic regions of the model in the space of the coupling parameter vs. energy, making them a useful guide in the study of the thermalization and quenched dynamics of the Dicke model. The results, particularly   the off-resonant cases, allow to establish firmly that, even if some relation between the onset of hard chaos and the ESQPT can be observed (noticeably in the resonant case), both phenomena, ESQPT and chaos, respond to different mechanisms and appear independently one of the other.     

The article is organized as follows. In section \ref{sec2} the semiclassical approximation of the Dicke model and the Hamiltonian describing the classical dynamics are presented, as well as its symmetries, integrable limits and how to build the Poincar\'e sections and estimate Lyapunov exponents. These exponents are used to quantify the percentage of chaos in the available phase space for a given energy.  In section \ref{sec4}   complete maps of chaoticity are shown in the energy-coupling space, for different sets of boson and qubits frequencies, paying special attention on the relation between the onset of chaos and the critical phenomena of the model. Finally, we expose our conclusions. The Appendices contain additional information about the semiclassical Hamiltonian, its equations of motion, its tangent space and  how to obtain the Lyapunov exponents. 

\section{ Dicke Hamiltonian and its classical limit}
\label{sec2}

\subsection{The Dicke Hamiltonian}

The Dicke model \cite{Dicke54}  has been widely employed to describe atom-field systems. It combines its simplicity with a rich variety of interesting features, like superradiance,  chaos, thermal and quantum phase transitions \cite{Emary03,Vid06,Lam05}.
The Hamiltonian has three terms
\begin{equation}
\begin{split}
H_{D}&=\omega a^{\dagger}a+\omega_{0}J_{z}+\\
&+\frac{\gamma}{\sqrt{\mathcal{N}}} \left(J_{+}+J_{-}\right)\left(a^\dagger +a \right).
\end{split}
\end{equation}
The first term is associated with a monochromatic quantized radiation field which has frequency $\omega$ and number operator $a^{\dagger}a$. The second term refers to a set of two-level atoms 
with excitation energy $\omega_{0}$. The number of excited atoms is accounted for by the third projection of a pseudo-spin collective operators $J_{z}$,  which together with $J_{+}$ and $J_{-}$ close an SU(2) algebra. The eigenvalues of $\mathbf{J}^{2}$ are $j(j+1)$, and the symmetric atomic subspace with  $j=\mathcal{N}/2$ includes the ground state. The ground state of the system exhibits a quantum phase transition when the atom-field interaction reaches the critical value $\gamma_{c}=\sqrt{\omega\omega_{0}}/2$. For smaller values of $\gamma$ it has no photons and no excited atoms, while at this atom-photon strength it becomes superradiant, suddenly  the number of photons and excited atoms becomes comparable to the total number of atoms in the system.

The mean-field description allows to capture many relevant aspects of the model. Critical exponents have been obtained for different observables  \cite{Emary03,Vid06,Chen0809}, and the presence of singularities around the QPT has been analyzed  \cite{OCasta11a,OCasta11,Hir13}.  
Another important feature of the Hamiltonian energy spectra is the presence of  the ESQPTs \cite{Bran13,Bas14A}, manifested as a singularity in the level density, order parameters, and wave function properties \cite{Cej06,Per11,Cap08,Str14}. Both the QPT and the ESQPT have been associated with the presence of classical chaos, and its quantum counterpart.


\subsection{The classical Hamiltonian}

Taking advantage of the algebraic structure of the hamiltonian, it is direct to build a semiclassical Hamiltonian and to obtain from it the semiclassical dynamics \cite{Bak13,SCcoherent1,SCcoherent2}. 
To this end we employ Glauber and Bloch coherent states, defined as follows:
\begin{equation}
\begin{split}
|\alpha\rangle&=e^{-|\alpha|^2/2}e^{\alpha a^\dagger}|0\rangle,\\
|z\rangle&=\frac{1}{\left(1+\left|z\right|^{2}\right)^{j}} e^{z J_+}|j, -j\rangle.
\end{split}
\label{ec:cs}
\end{equation}
The classical Hamiltonian is calculated as the expectation value of the Hamiltonian operator in the coherent state product \cite{MAM92}.  The coherent states are built as functions of the complex variables $\alpha$ and $z$. From them, the canonical variables $(p,q)$ and $(j_z,\phi)$ are defined as  $\alpha=\sqrt{\frac{j}{2}}(q+i p)$ with $q$ and $p$ real values for the photonic sector, and the stereographic projection of $z=\tan(\theta/2)e^{i \phi }$, with $\tilde{j_z}=(j_z/j)=-\cos\theta$ and $\phi=\arctan(j_y/j_x)$, where $\theta$ and $\phi$ are spherical angular variables of a classical vector $\vec{j}=(j_x,j_y,j_z)$  ($|\vec{j}|=j$), with $\theta$ measured respect to the negative $z$-axis. 

The classical Hamiltonian per particle (see Ref. \cite{Bas16}), expressed in terms of these canonical variables, reads 
\begin{equation}
\begin{split}
&h_{cl}(p,q,\tilde{j_z},\phi)=\frac{\langle \alpha, z| H_D|\alpha, z\rangle}{j}=\\
&=\omega_{0}\tilde{j_{z}}+\frac{\omega}{2}\left(q^{2}+p^{2}\right)+2\gamma \sqrt{1- \tilde{j_{z}}^{2} }\,q\,\cos \phi.
\end{split}
\label{ec:hacl}
\end{equation}
As discussed below, the only integrable limits of the model are $\gamma=0$, $\omega_0=0$ or $\omega=0$.  For every other value of the Hamiltonian parameters, the semiclassical version of the Dicke Hamiltonian is non-integrable. This is the main subject of this work.   

The classical energy surface, defined by  $h_{cl}(p,q,\tilde{j_z},\phi)= \epsilon$, is depicted in Fig. \ref{fig:SupEn_e_-1,4_g_2} for three different energies  $\epsilon=-1.4\,\omega_0,-1.1 \omega_0$ and  $-0.5 \omega_0$. The changes in their topology reflect both the QPT and the ESQPT.
\begin{figure*}
\centering
\begin{tabular}{c c}
(a)&(b)\\
\includegraphics[width= 0.4 \textwidth]{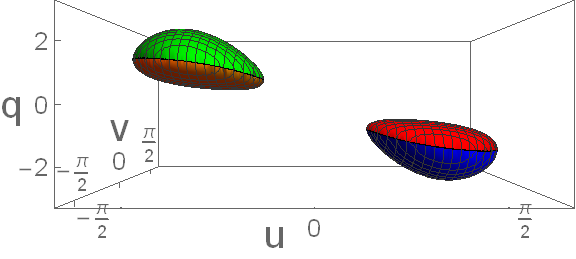}& \includegraphics[width= 0.4 \textwidth]{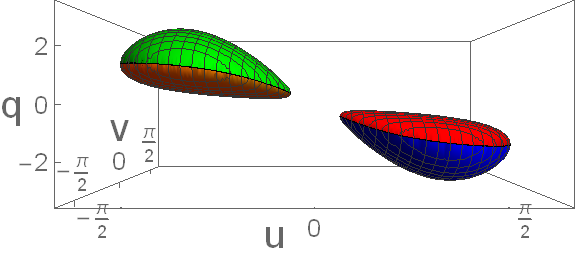}\\
(c) &  \\
\includegraphics[width= 0.4 \textwidth]{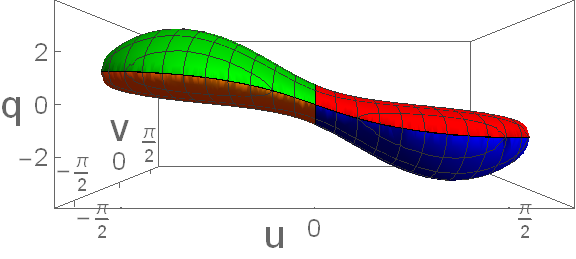} & \includegraphics[width= 0.1 \textwidth]{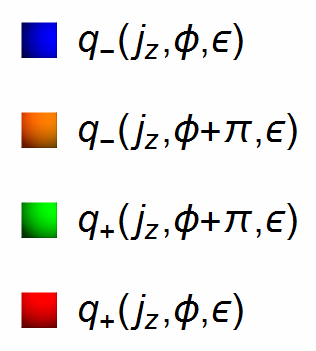}\\
\end{tabular} 
\caption{(Color online)
Energy surface shell for $\gamma=2.0\,\gamma_{c}$, in coordinates $(u=\theta\cos\phi,v=\theta\sin\phi,q)$. Three energies are shown: a) $\epsilon=-1.4\,\omega_0$,  b) $\epsilon=-1.1 \omega_0$ and c) $\epsilon=-0.5 \omega_0$.}
 \label{fig:SupEn_e_-1,4_g_2}
\end{figure*}

The semiclassical ground state energy $\epsilon_{0}(\gamma)$ has the form \cite{OCasta09,OCasta11,Nah13} 
 \begin{equation}
\epsilon_{0}(\gamma)=\left\{\begin{array}{lr} -\omega_0  & {\hbox{for }} \gamma\leq \gamma_c,  \\
-\frac{\omega_0 }{2} \left(\frac{\gamma_c^2}{\gamma^2}+ \frac{\gamma^2}{\gamma_c^2}\right) &{\hbox{for }} \gamma > \gamma_c . \\ \end{array} \right. 
\label{eq:gse}
\end{equation}
where the QPT can be observed, manifested as a discontinuity on its second derivative. 

The energy spectra is only lower bounded. 
As the energy increases,  the available phase space surfaces acquire  different structures. 
The changes in the form of the energy surfaces signal the ESQPT. Different sectors of the Bloch sphere are available in the normal and superradiant phases.Their accesible volume can be quantified using the DoS, which displays singularities at the ESQPT  \cite{Bas14A,Str14,Bran13,Bas15}.


\subsection{Symmetries in the Hamiltonian}

The quantum version has a discrete symmetry coming from the fact that the Hamiltonian does not mix states with different parity (even or odd) number of excitations. The parity operator associated is $P=e^{i\pi(J_z+j+a^\dagger a)}$ with eigenvalues $\Pi=\pm 1$. This symmetry is reflected in the classical version as the invariance of the Hamiltonian under the transformation
\begin{equation}
(\phi,q)\rightarrow (\phi+\pi,-q),
\label{ec:parClass}
\end{equation} 
which helps to simplify the numerical efforts in the study of the classical dynamics. The QPT is associated with the spontaneous breaking of this symmetry \cite{Puebla13}. In the superradiant region, with $\gamma>\gamma_c$, the low energy classical trajectories (including the minimal energy fixed points) are two fold degenerate, i.e., there exist two different trajectories which can be obtained one from the other by the parity transformation (\ref{ec:parClass}). Moving up in energy the parity symmetry is spontaneously restored crossing the ESQPT, at $\epsilon \geq -\omega_o$, where the trajectories become invariant under the parity transformation.

In Fig.\ref{fig:SupEn_e_-1,4_g_2}, the spontaneous breaking and restoration of the parity symmetry is illustrated by showing the energy shell for $\gamma=2\gamma_c$   (with $p=0$),  corresponding to three different energies  $\epsilon=-1.4 \omega_o, -1.1 \omega_o$ and $-0.5\omega_o$, the first two below the ESQPT and the last one above.  
We will use this kind of surfaces (with $p=0$) to explore the dynamics of the system, employing them to obtain Poincar\'e sections and calculating the Lyapunov exponents for a large sample of points over these surfaces, as explained in the next section. The surfaces are obtained, 
for  given energy $\epsilon$, by selecting $p=0$. The values of the variable $q$ are calculated solving the quadratic equation $h(p=0,q,\tilde{j_z},\phi)=\epsilon$ which gives two different values of $q$,

\begin{equation}
\begin{split}
&q_{\pm}(j_z,\phi,\epsilon)=-\frac{2\gamma}{\omega}\sqrt{1-\tilde{j_{z}}^{2}}\,\cos\phi\,+\\
&\pm\sqrt{\frac{4\gamma^{2}}{\omega^{2}} \left(1-\tilde{j_{z}}^{2}\right)\,\cos^{2}\phi+\frac{2}{\omega}\left(\epsilon-\omega_{0}\tilde{j_z}\right)}.
\end{split}
\label{ec:qpm}
\end{equation}
The allowed values of the remaining variables, $\phi$ and $j_z$, for a given energy, are explicitly given in \cite{Bas14A}. 


\subsection{Integrable limits of the classical Dicke model}

As mentioned  before, the Hamiltonian (\ref{ec:hacl}) is integrable for a zero value in any of the three parameters of the model. 

For  zero coupling ($\gamma=0$),  the Hamiltonian becomes independent on the variable $\phi$ \textit{i.e.} $\dfrac{\partial H}{\partial \phi}=0$, making   $j_z$  a constant of motion.  

In the case $\omega=0$, the Hamiltonian becomes independent on the variable $p$, implying that $\dot{q}=0$ and $q$ becomes  a constant of motion.

When  $\omega_0=0$, a canonical transformation $(j_z,\phi)\rightarrow(j'_x,\phi_x)$, with $\phi_x$ the  azimuth angle in the plane $z$-$y$, can be performed. The new Hamiltonian reads
\begin{equation}
h=\frac{\omega}{2}\left(q^{2}+p^{2}\right)+2\gamma \,q\,\tilde{j'_{x}},
\end{equation}
independent of the angular variable $\phi_x$, making $j'_x$
a constant of motion. The Hamiltonian is, thus, equivalent to a displaced harmonic oscillator 
\begin{equation}
q'^{2}+p^{2}=\frac{2H}{\omega}+\frac{4\gamma^{2}j'^{2}_x}{\omega^{2}}
\end{equation}
where  $q'=q+\dfrac{2\gamma\, j'_x}{\omega}$. 
 
In all the previous cases, the Hamiltonian is effectively reduced to a conservative one-dimensional system $H=H(p,q)$ or $H=H(\tilde{j_z},\phi)$,  which is always integrable and unable to present chaotic dynamics. Except for these particular limiting cases, the system is not integrable, making room for chaotic behavior.


\subsection{Poincar\'e sections and Lyapunov exponents}

While the canonical variables employed above are useful in the analysis of the symmetries and the integrable limits of Hamiltonian (\ref{ec:hacl}), the numerical integration of the equations of motion is more stable employing variables which are bounded in the phase space. To this end we introduce new canonical variables in the atomic sector  $\tilde{P}=-j\sqrt{2(1+\tilde{j_z})}\sin\phi$ and $Q=\sqrt{2(1+\tilde{j_z})}\cos\phi$, which satisfy $\{\tilde{P},Q\}=-1$. Defining $P=\tilde{P}/j$, the classical Hamiltonian reads
\begin{equation}
\begin{split}
h_{cl}=\frac{\omega_0}{2}\left(Q^{2}+P^{2}\right)+\frac{\omega}{2}\left(q^{2}+p^{2}\right)+ \\
2\gamma\,q\,Q \sqrt{1- \frac{1}{4}\left(Q^{2}+P^{2}\right)}-\omega_0.
\end{split}
\label{ec:haclPQ}
\end{equation}

Building up from the analysis presented in Ref. \cite{Bas16},  in this section we present the Poincar\'e sections for some representative Hamiltonian parameters and excitation energies. They help to visualize in a familiar form the presence of regular and chaotic orbits. We compare them with the maximal Lyapunov exponents \cite{Lyapunov92,Oseledets68,Benettin80}, which allow to transit form a qualitative to a quantitative description of chaos. 
A detailed description of the mathematical procedure to obtain the equations of motions and the Lyapunov exponents is given in Appendix \ref{Ap:LCE}. The classical trajectories are obtained by numerical integration of the equations of motion, Eqs. (\ref{FdF1}) and (\ref{eq:Din_PQ}).

In Fig. \ref{im:Tray-1,4g2}, the extreme sensitivity to initial conditions is illustrated by showing two trajectories starting with very close initial conditions. They begin to separate for $t\sim 50$, having at larger times clear different behaviors.  When the divergence between the trajectories is exponential in the tangent space of the respective phase space, the largest Lyapunov exponent associated with this specific point in the phase space (the initial condition) is positive,  unveiling a chaotic nature.
\begin{figure}
 \centering
\begin{tabular}{c c}
\includegraphics[width= 0.2 \textwidth]{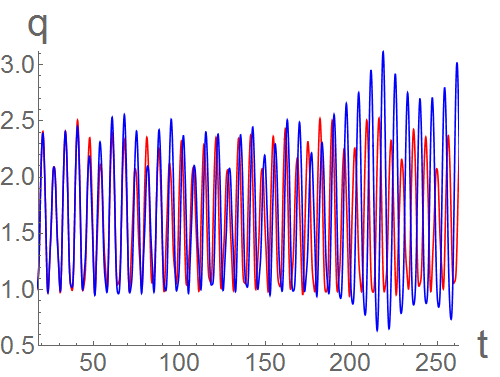} & \includegraphics[width= 0.2 \textwidth]{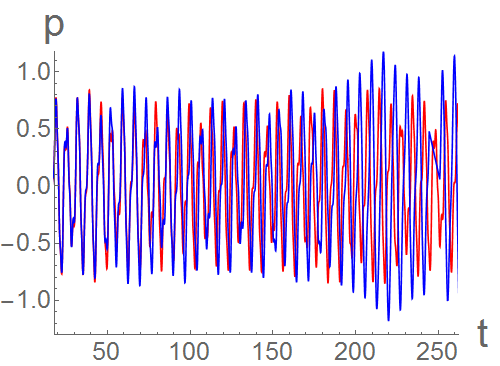}  \\ 
(a) & (b) \\
\includegraphics[width= 0.2 \textwidth]{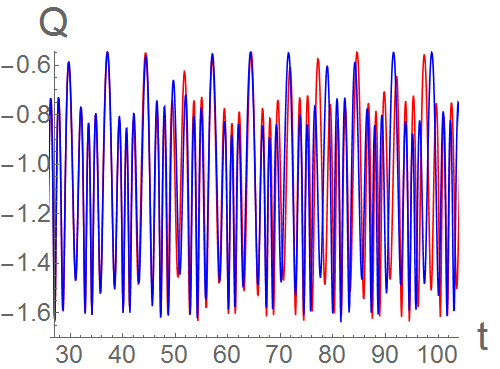} & \includegraphics[width= 0.2 \textwidth]{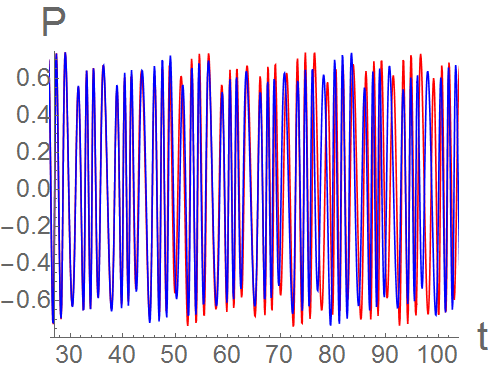} \\ 
(c) & (d)  \\
\end{tabular} 
  \caption{\small (Color online) Solutions of the canonical variables versus time for two close trajectories each, for $\epsilon=-1.4\,\omega_0$, $\gamma=2\gamma_{c}$, with initial condition $(p_0,q_0,P_0,Q_0)=(0,q_+(\epsilon),0.648,1.371$)}. 
   \label{im:Tray-1,4g2}               %
\end{figure}

In order to have as many different trajectories as allowed for a given set of parameters, we restrict the analysis to
the plane $p=0$, which can be expressed in terms of the variables $Q$ and $P$ through the energy conservation $h_{cl}(p=0,q,P,Q)=\epsilon$. 
The intersection of the classical orbits with this surface defines the Poincar\' e surface sections.

Figure \ref{im:en-1,4g2} a) displays the energy allowed regions as a function of atom-field coupling $\gamma$. The three dots represent three values of the energy excitation, $\epsilon=-1.4\,\omega_0$ (orange), $\epsilon=-1.1\,\omega_0$ (red) and $\epsilon=-0.5\,\omega_0$ (green), for $\gamma=2\,\gamma_{c}, \omega = \omega_0$ where we will concentrate the first part of the analysis. Fig.  \ref{im:en-1,4g2} b) shows, employing the same color code, the contours of each surface energy in the canonical variables $(Q,P)$.

\begin{figure}
 \centering
\begin{tabular}{c c}
\includegraphics[width= 0.18 \textwidth,height= 0.2 \textheight]{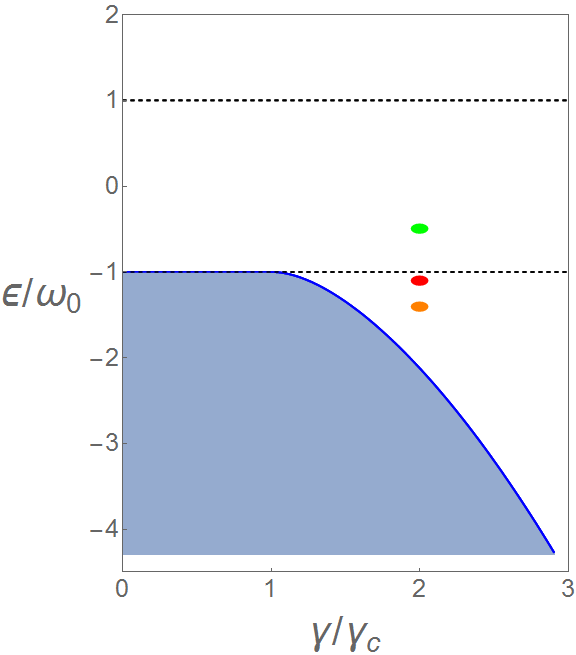} & 
\includegraphics[height= 0.2 \textheight]{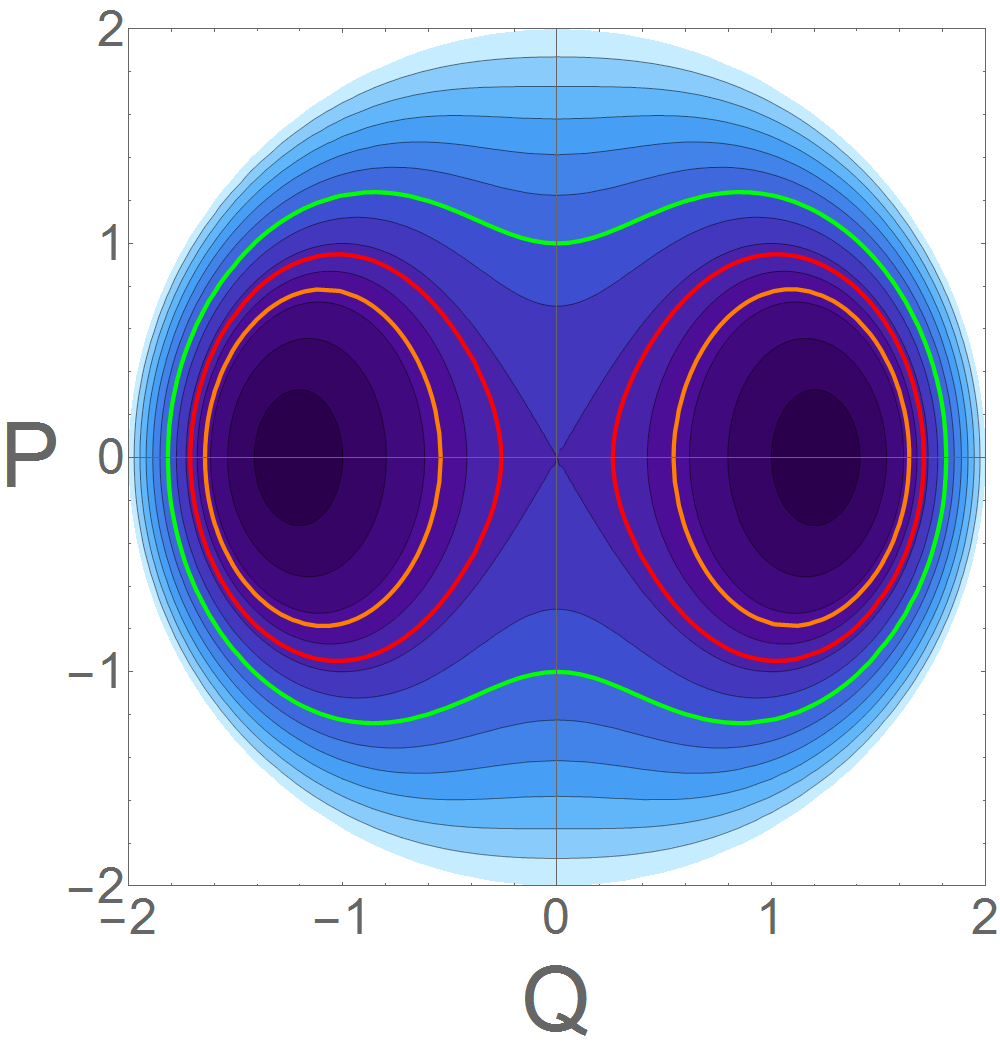}  \\ 
(a) & (b) \\
\end{tabular}
\caption{\small (Color online) Points of analysis (a) for $\gamma=2\,\gamma_{c}, \omega = \omega_0$ and $\epsilon=-1.4\,\omega_0$ (orange), $\epsilon=-1.1\,\omega_0$ (red), $\epsilon=-0.5\,\omega_0$ (green). The blue curve depicts the ground state energy as a function of the coupling parameter $\gamma$. (b) Contours of the surface energy $\epsilon$ in the canonical variables ($Q,P$), for coupling value $\gamma=2\,\gamma_{c}$ and $\epsilon=-1.4\,\omega_0$ (orange), $\epsilon=-1.1\,\omega_0$ (red), $\epsilon=-0.5\,\omega_0$ (green). }
\label{im:en-1,4g2}               
\end{figure}

\begin{figure*}
\centering
\begin{tabular}{c c c }
\includegraphics[width= 0.25 \textwidth]{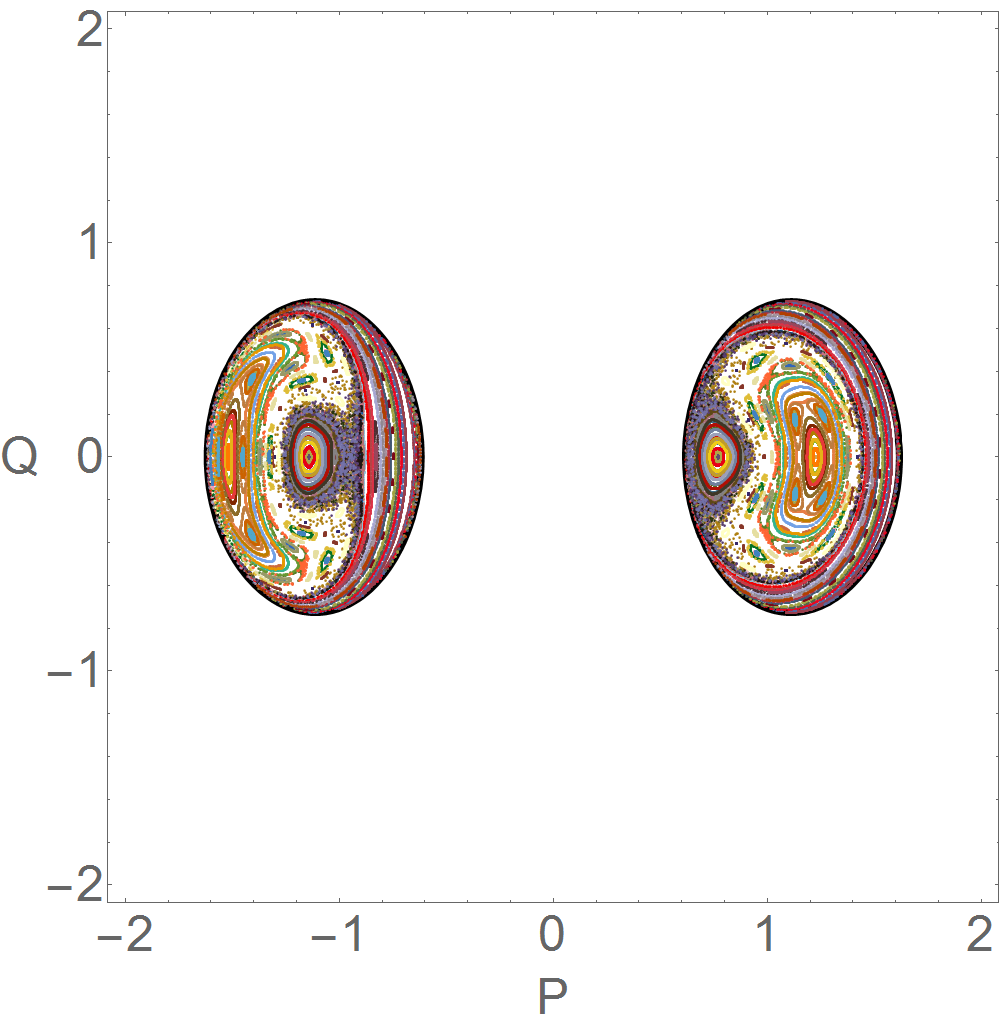} & \includegraphics[width= 0.25 \textwidth]{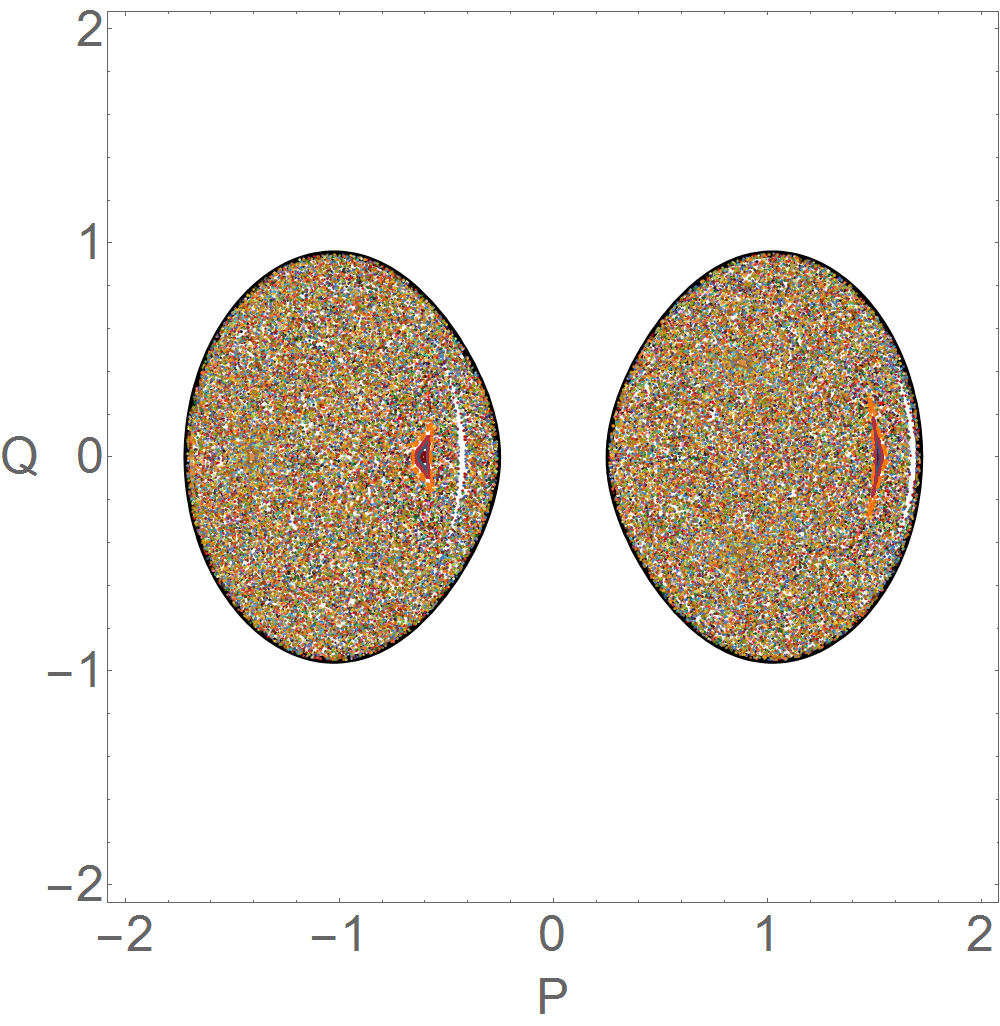} & \includegraphics[width= 0.25 \textwidth]{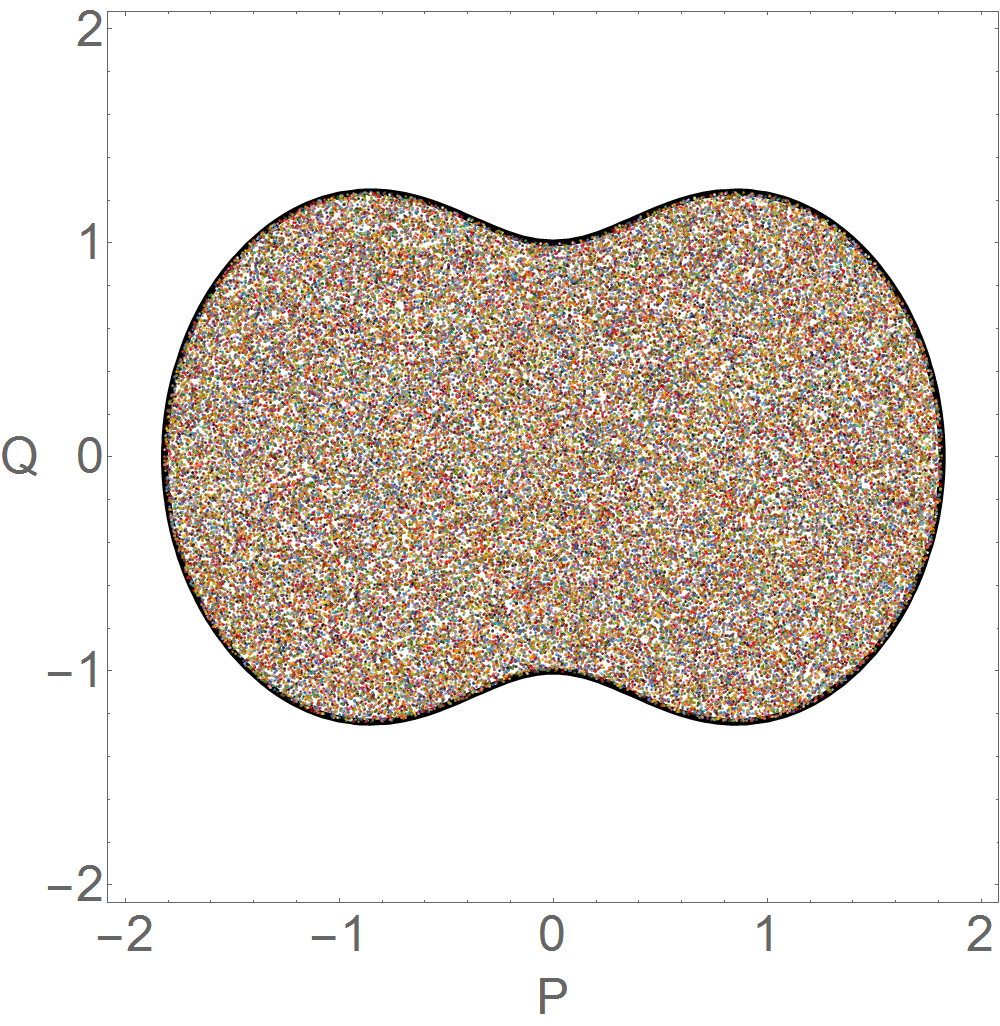} \\ 
\includegraphics[width= 0.25 \textwidth]{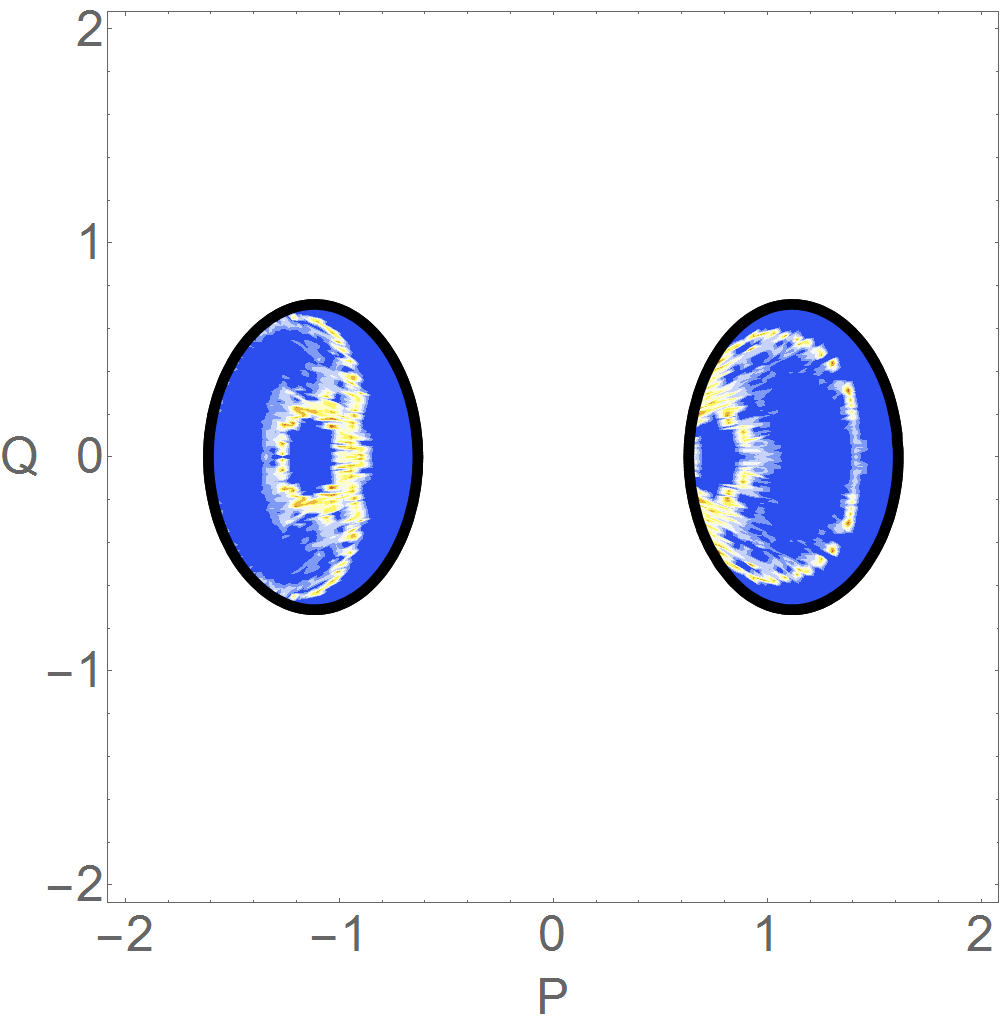} & \includegraphics[width= 0.25 \textwidth]{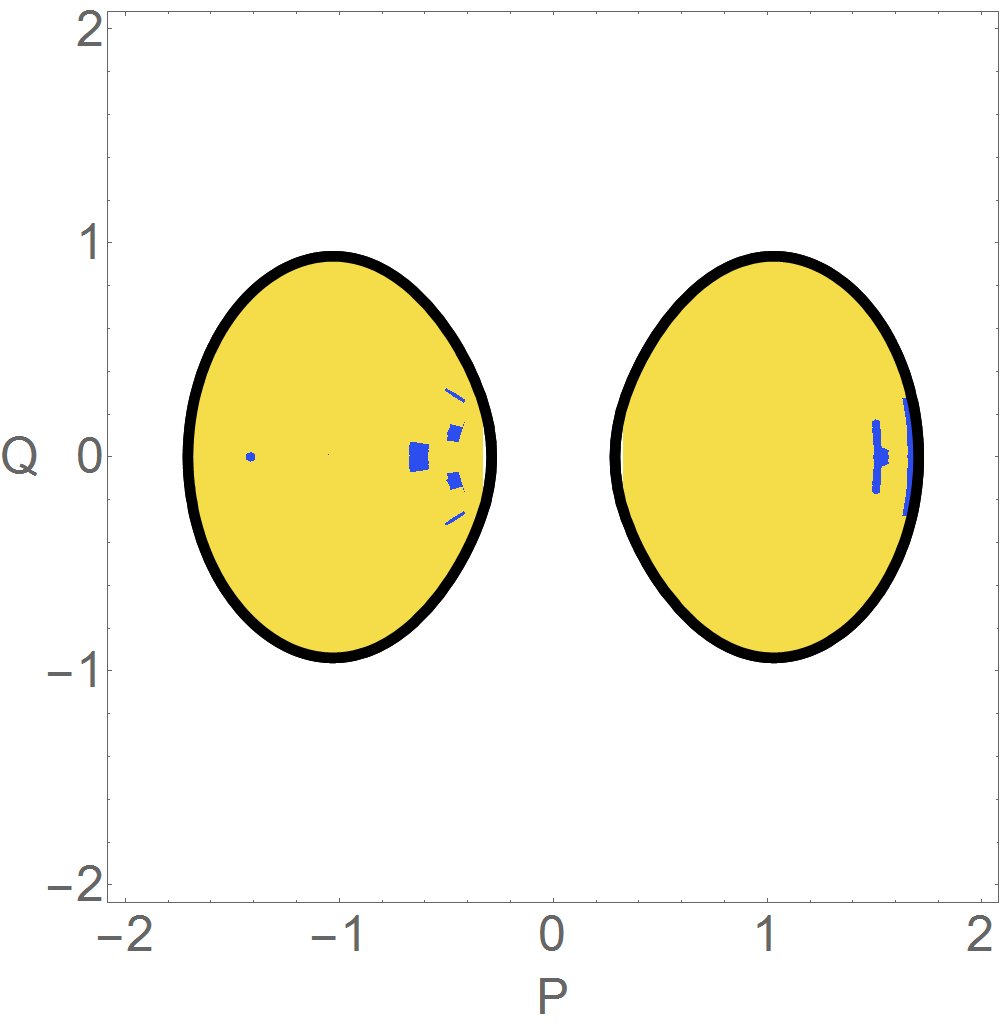} &  \includegraphics[width= 0.25 \textwidth]{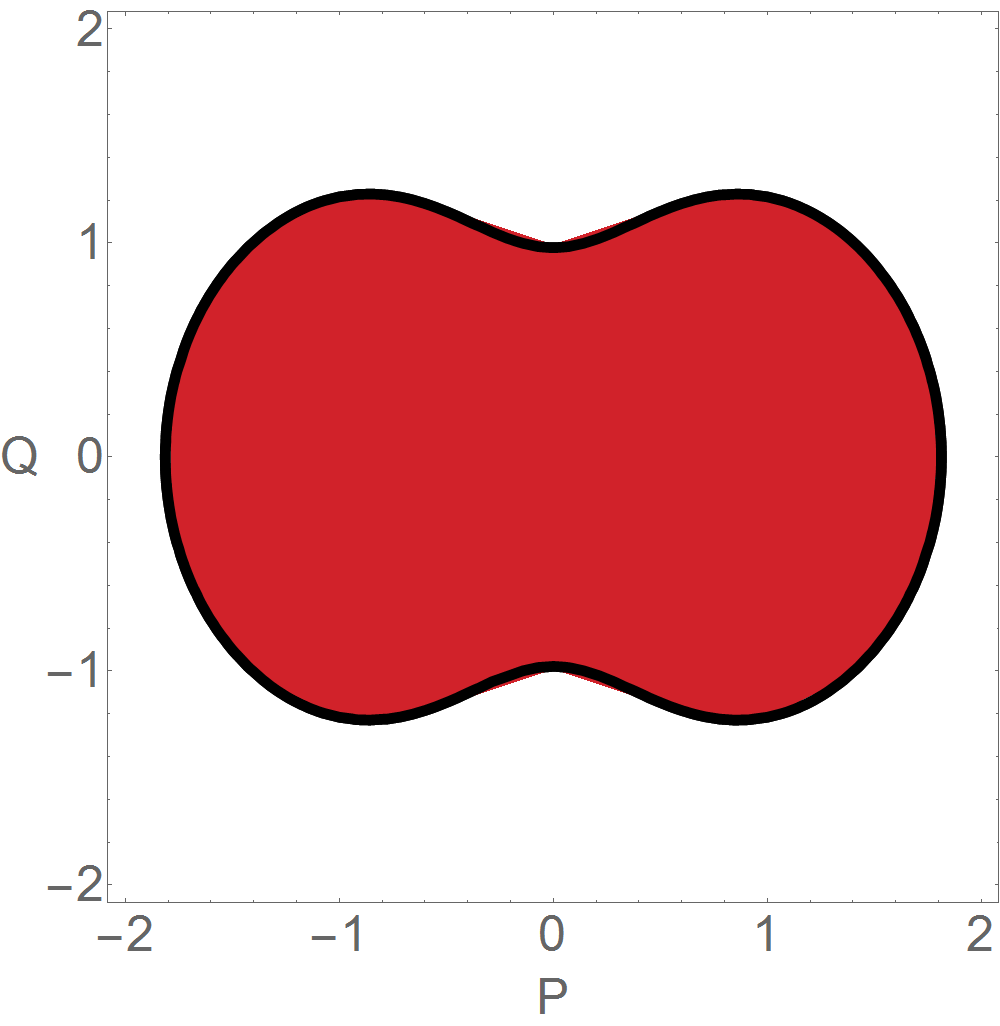} \\ 
 \includegraphics[width= 0.25 \textwidth]{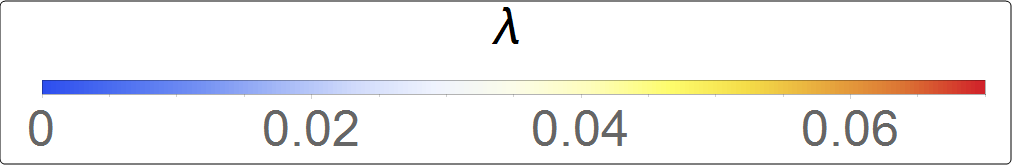}  & \includegraphics[width= 0.25 \textwidth]{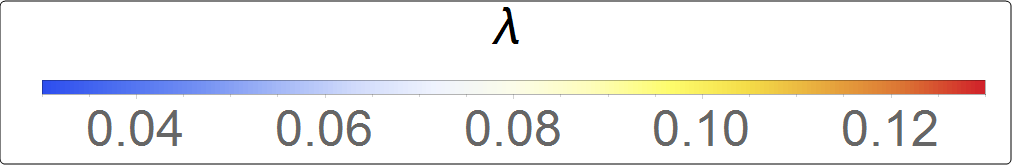}   &  \includegraphics[width= 0.25 \textwidth]{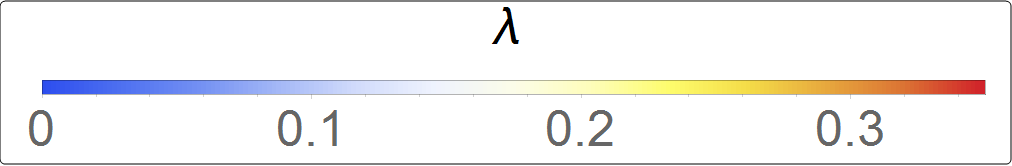}   \\
(a) & (b) & (c)  \\
\end{tabular} 
\caption{ (Color online) Poincar\'e sections (top) and Lyapunov exponents (bottom) in the proyection of variables $(Q,\,P)$, for $\gamma= 2 \,\gamma_c$ and $\epsilon= -1.4 \,\omega_0$ (a), $\epsilon= -1.1 \,\omega_0$ (b), $\epsilon= -0.5 \,\omega_0$ (c). In the Poincar\'e sections the colors are associated with different classical trajectories. For the Lyapunov exponents the color code is given on the bar (down). Blue depicts the regular regions. }
\label{fig:poinc}
\end{figure*}

In Fig. \ref{fig:poinc} we present  Poincar\'e sections and the Lyapunov exponents for the same energies $\epsilon= -1.4 \,\omega_0$, $-1.1 \,\omega_0$ and $-0.5 \,\omega_0$, for   $\gamma= 2\,\gamma_c$ in resonance ($\omega=\omega_0$), as functions of $Q$ and $P$. In all cases, the regions with scattered  points in the Poincar\'e sections have a Lyapunov exponent different to zero, allowing to quantify the presence of chaos in the system, which is qualitatively suggested by the Poincar\'e sections.  For $\epsilon= -1.4\,\omega_0$ regular and chaotic regions coexist, as energy increases the stability islands shrinks, at the largest energy shown they have completely disappeared and each trajectory explores all the available phase space. 

\begin{figure}
 \centering
        \includegraphics[width= 0.45 \textwidth]{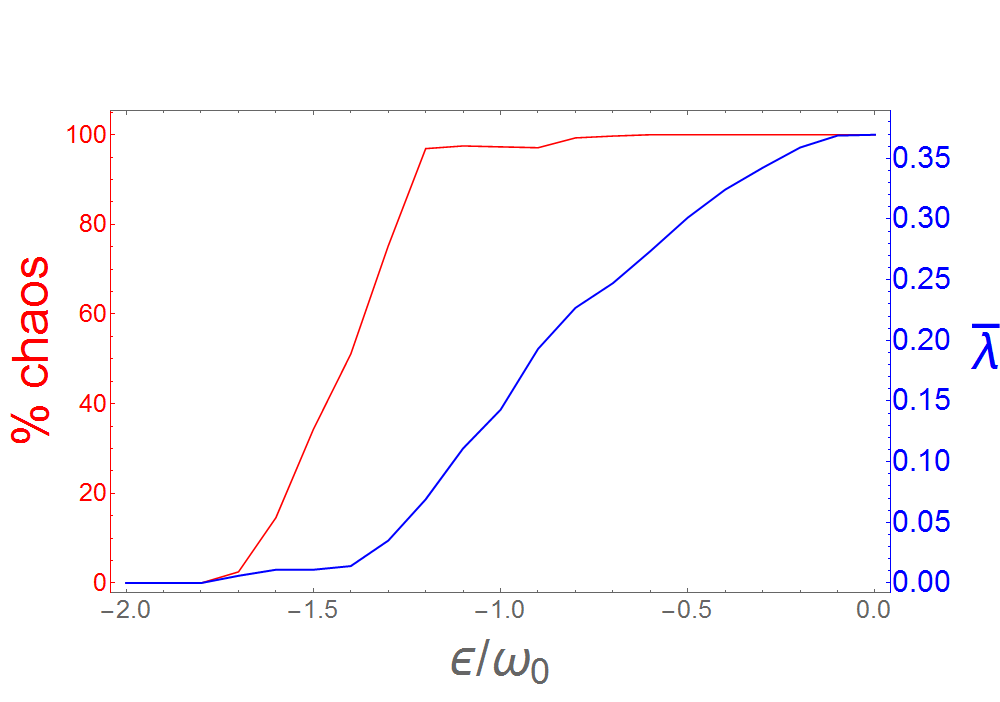}
  \caption{\small (Color online) Percentage of chaos and average Lyapunov exponent in phase space with coupling $\gamma=2\gamma_{c}$ in resonance.}
 \label{fig:Porc_caos_g_2}
\end{figure}

With the  Lyapunov exponents calculated for a large sample of points uniformly distributed in the available phase space (see Appendix \ref{Ap:LCE_Sample}), we are able to estimate the percentage of chaos for a given energy by taking the ratio of the number of points with  a Lyapunov exponent different from zero (in practice larger than $\lambda_{min}=0.002$) to the total number of points in the sample. Likewise,  we can calculate  the mean value of the Lyapunov exponent for a given energy. Both measures reflect the onset of chaos in the system as a function of energy, quantifying the transition from a regular regime at low energies  to an ergodic regime at larger ones. Results for the case   $\gamma=2\,\gamma_{c}$ in resonance $\omega=\omega_0$ are shown in Fig. \ref{fig:Porc_caos_g_2}. 

As expected, for values close to the minimum energy $\epsilon_{0}$ the percentage of chaoticity is null and remains close to zero up to $\epsilon\sim -1.7 \omega_0$, from there it increases with the energy, attaining the saturation value at energy $\epsilon\sim -1.2 \omega_0$.
The average value of the Lyapunov exponent presents a similar behavior, at energies lower than $\epsilon\sim -1.2 \omega_0$ is very close to zero, and from there  increases monotonically.

\begin{figure*}
  \centering
  \begin{tabular}{c c c }
\multicolumn{3}{c}{\includegraphics[width= 0.4 \textwidth]{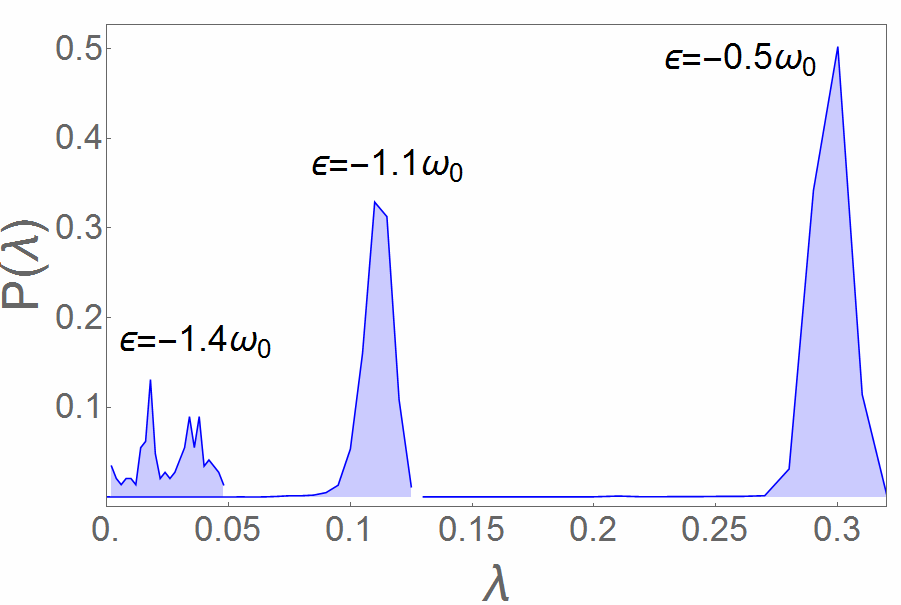}} \includegraphics[width= 0.08 \textwidth]{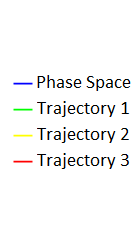} \\
  \includegraphics[width= 0.3 \textwidth]{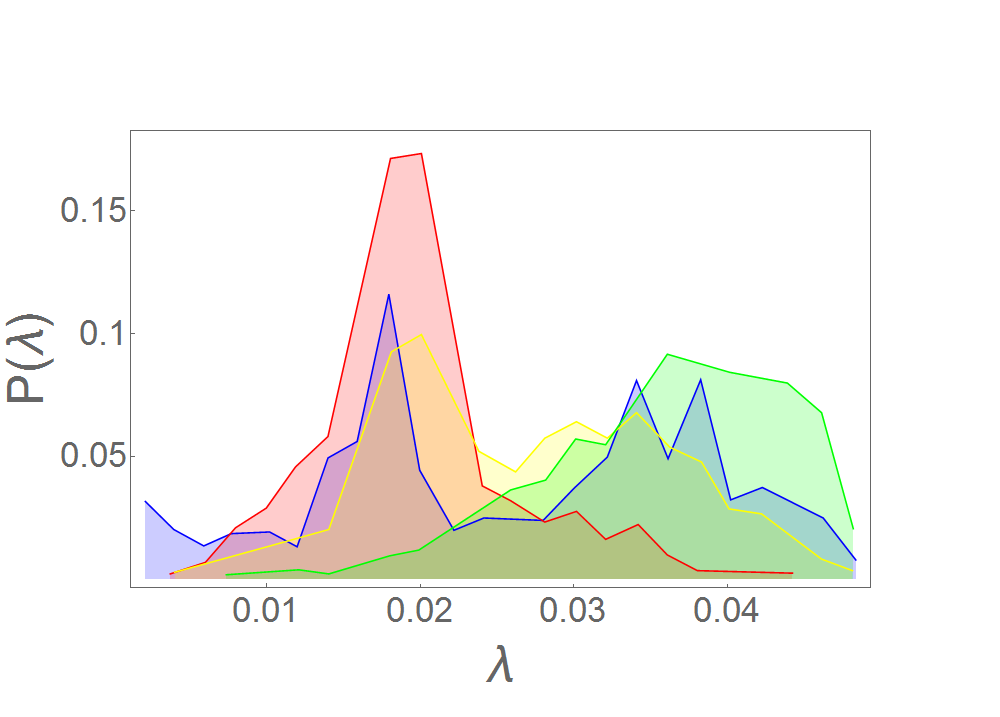} & \includegraphics[width= 0.3 \textwidth]{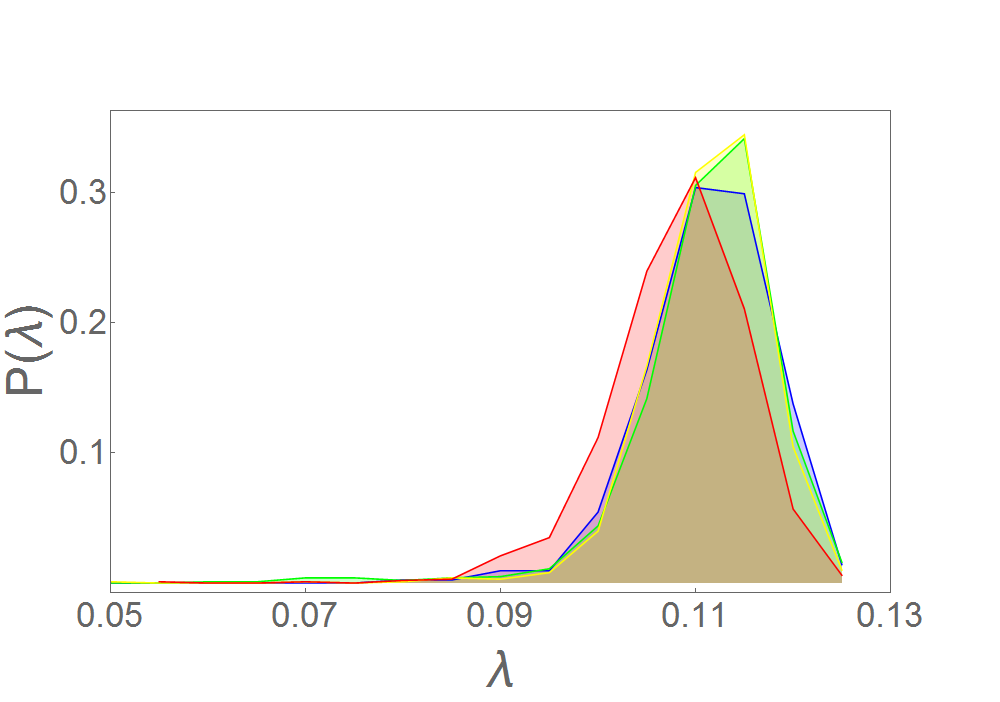} & \includegraphics[width= 0.3 \textwidth]{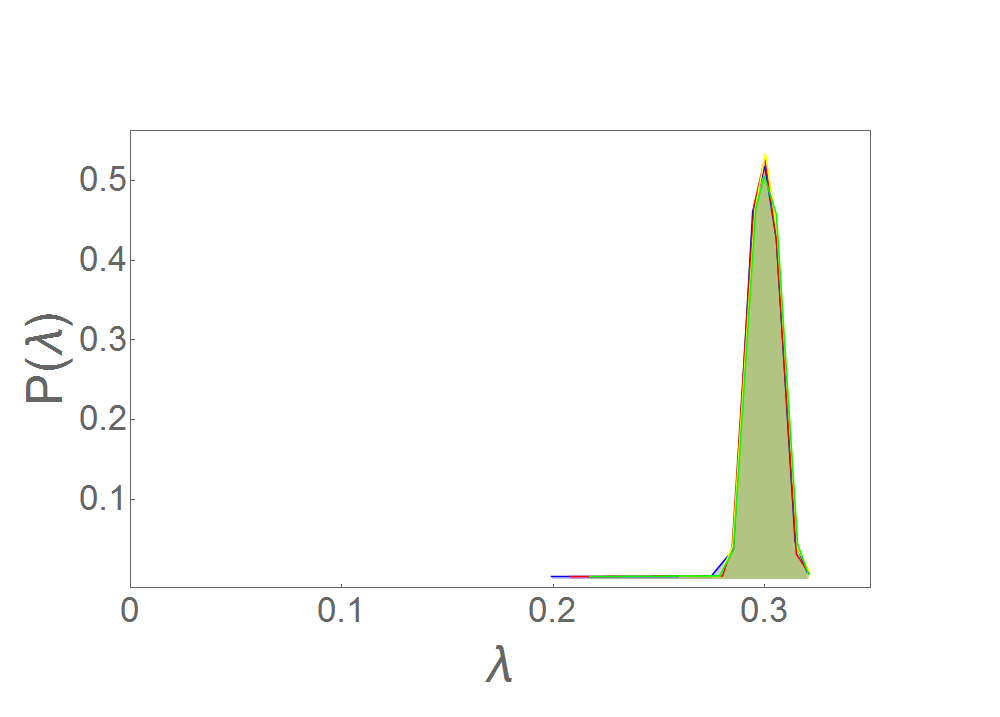} \\
 (a) & (b) & (c)
   \end{tabular}
 \caption{\small  (Color online) Probability distribution (top) for the LCE of three different surface energies in the respective phase space, (bottom) distribution of probability for three trajectories with initial condition $\bar{x}_1(0)$ $=(p_o,q_o,P_o,Q_o)=(0, q(\epsilon), 0, 0.948)$, $\bar{x}_2(0)=(0, q(\epsilon), 0, 0.707)$, $\bar{x}_3(0)=(0, q(\epsilon), 0, 0.547)$ for $\epsilon=-1.4\,\omega_0$ (a), $\epsilon=-1.1\,\omega_0$ (b) and $\epsilon=-0.5\,\omega_0$ (c).}
   \label{fig:dis_lyap}                
\end{figure*}


\subsection{Distributions of the Lyapunov exponents}

The Lyapunov exponent, as defined in Eqs. (\ref{ec:LCE}) and  (\ref{ec:IC}), depends on the initial condition $\textbf{x}(0)$ in phase space. If we follow a trajectory and recalculate the Lyapunov exponent considering as initial point the value of the coordinates at time $t$, we will not obtain in general the same value unless the system is fully ergodic \cite{Cencini10,Cornfeld82}.  In this way we can associate a Lyapunov exponent to each point in phase space, and to the set of points along a given trajectory.  

The distributions of the Lyapunov exponents in the available phase space is presented in Fig. \ref{fig:dis_lyap} for the three  representative energies mentioned above.  In the case of the two largest energies, the distributions are unimodal and their variances are very small. For  the higher energy, $\epsilon=-0.5\,\omega_0$,  the percentage of chaos is very close to 100 percent and the mean Lyapunov exponent  is $  \overline{\lambda}= 0.301$. For the intermediate energy  $\epsilon=-1.1\,\omega_0$, small islands of stability are present, giving a  percentage of chaoticity close to  97.5\%, and  an   average Lyapunov exponent $\overline{\lambda}\sim 0.111$. These results are  consistent with  the fact that the available phase space is ergodic or very close to that, as can be seen in Fig. 4. Consequently,  a  very similar  Lyapunov exponent is obtained whatever point (with its respective infinitesimal neighborhood) in the phase space is considered. Finally, for the lowest energy shown,  $\epsilon=-1.4\,\omega_0$, the phase space is mixed and contains  several regular and chaotic regions. As a  consequence the distribution of Lyapunov exponents  presents various local maxima, including a maximum at $\lambda=0$,  associated with  regular trajectories.  The   chaotic regions   cover   51.1\%  of the available phase space, and the average Lyapunov exponent  is 0.014.  

The above mentioned ergodicity at large energies can also be seen at the distribution of the Lyapunov exponent on  particular trajectories. At the bottom of Fig. \ref{fig:dis_lyap} the  distributions of the Lyapunov exponents are shown for three different trajectories at each energy. Whereas for the two largest energies,  the distributions over each trajectory and over the available phase space are almost identical, in the low energy, mixed case, the distributions vary drastically from one trajectory to another, and none is equal to the distribution of the Lyapunov exponents over the available phase space. 

  
\section{Chaos maps in  space $\gamma$ and $\epsilon$}
\label{sec4}
\begin{figure*}
 \centering
  \begin{tabular}{c c}
    \includegraphics[width= 0.45 \textwidth]{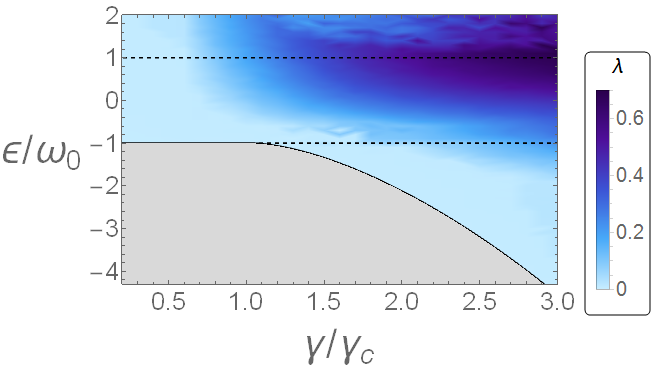} & \includegraphics[width= 0.5 \textwidth]{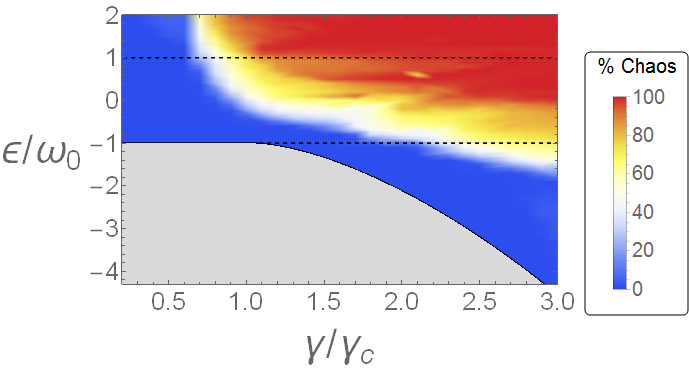}\\ 
    (a) & (b)\\
    \includegraphics[width= 0.45 \textwidth]{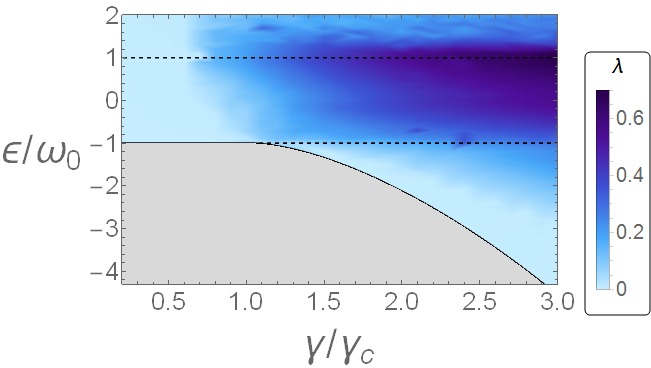} & \includegraphics[width= 0.5 \textwidth]{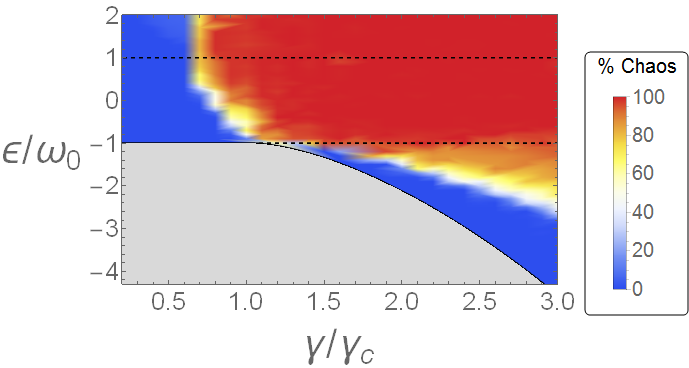}\\ 
    (c) & (d)\\
    \includegraphics[width= 0.45 \textwidth]{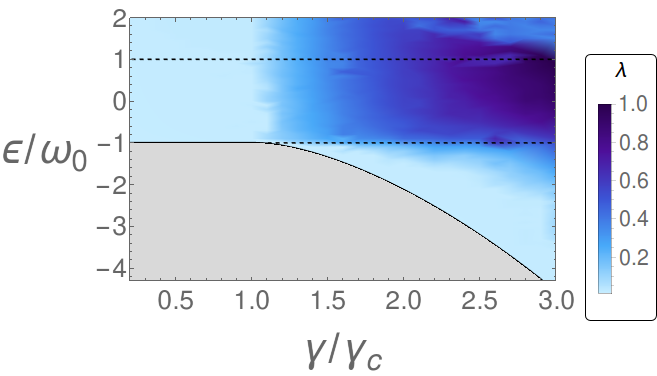} & \includegraphics[width= 0.5 \textwidth]{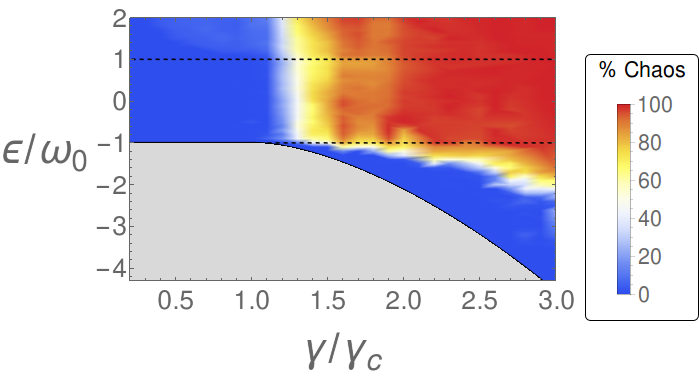}\\ 
    (e) & (f)\\
  \end{tabular}
 \caption{\small  (Color online) Mapping for Lyapunov exponent on energy surface (left), percentage of chaoticity (right) for diferents vales of $\omega$: $\omega=\omega_0/4$ (top), $\omega=\omega_0$ (center), $\omega=4\omega_0$ (bottom).
 The dotted lines represent the energies of the ESQPT.
 }
   \label{im:ZONAS_L_C}                
\end{figure*}

The analysis presented in the previous section for $\gamma=2\gamma_c$ in resonance $\omega_0=\omega$ is  extended in this section to   couplings in the interval $\gamma\in[0,3 \gamma_c]$ for  three different sets of qubits and boson frequencies $\omega=\omega_0/4$, $\omega=\omega_0$ and $\omega=4 \omega_0$. The percentage of chaos and the mean  Lyapunov exponents over the available phase space for a given energy were calculated as explained in the previous section. The result of these calculations are complete maps of chaos of the classical Dicke model, which are   presented in Fig. \ref{im:ZONAS_L_C}.

It is worth to mention that the computational demand of this exercise is huge. As explained in Appendix \ref{sampling}, to determine the Lyapunov exponent at a given point in phase space, an average over thousands of initial conditions in the vicinity of the selected point is performed.  To map the available phase space for a given set of Hamiltonian parameters at a given energy with the resolution presented in Fig. \ref{fig:dis_lyap}, the Lyapunov exponent is calculated for around thousand points. Over this phase space, the average Lyapunov exponent and the percentage of chaotic points is evaluated. This procedure is repeated for a large number of energy values $\epsilon$ and coupling constants $\gamma$, totaling about four thousand pairs $(\epsilon,\gamma)$ for each of the three maps presented in Fig.\ref{im:ZONAS_L_C}.

\subsection{Chaos and QPT}
\begin{figure*}
\centering
\begin{tabular}{c c c}
 \includegraphics[width= 0.4 \textwidth]{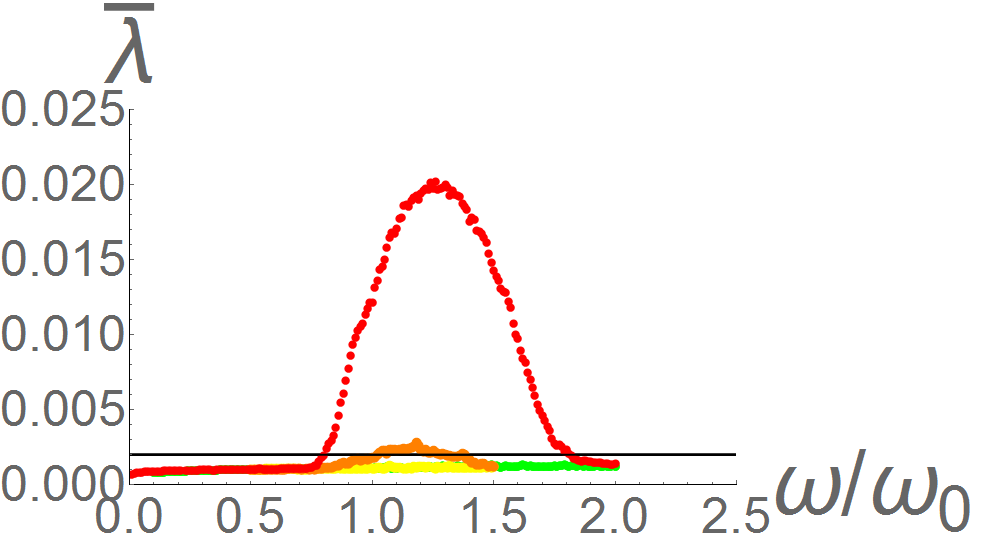} &\includegraphics[width= 0.4 \textwidth]{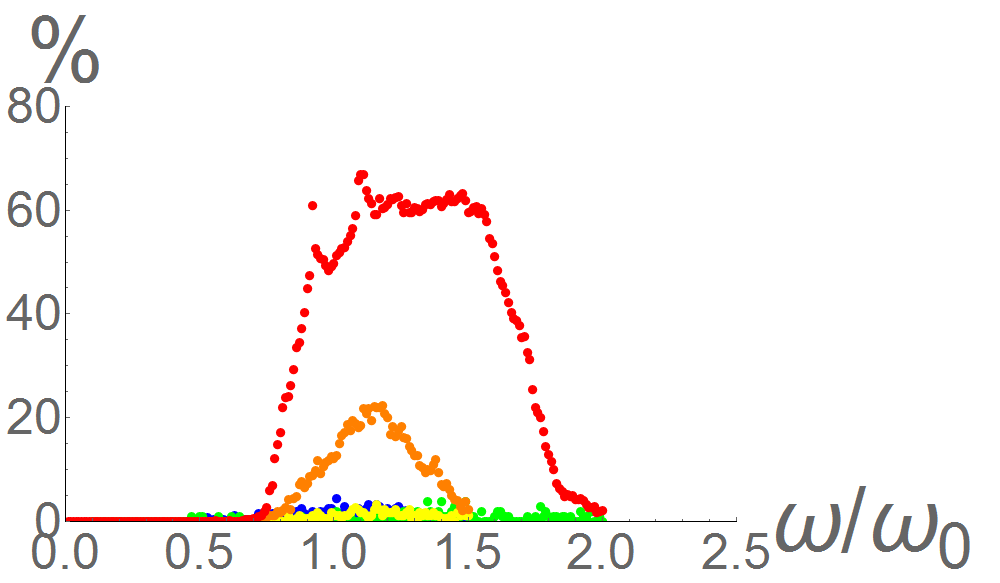} & \includegraphics[width= 0.1 \textwidth]{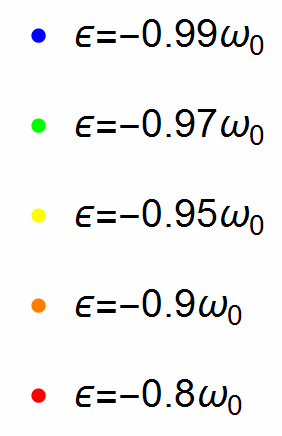} \\ 
(a) & (b) & \\
\end{tabular} 
\caption{\small  The average Lyapunov exponent  (left) and the percentage of chaoticity (right), plotted as functions of the photon energy $\omega$ for $\gamma=\gamma_c$ and different values of the excitation energy $\epsilon$ close to the ground state energy $\epsilon_0$.}
\label{fig:por_promL_c}
\end{figure*}

A common characteristics in the three maps is the existence of a regular regime, with zero mean Lyapunov and chaos percentage,  visible at low excitation energies for almost any coupling, and  also at high energies for couplings close to zero. As mentioned above, these results are expected because $\gamma=0$ is one of the integrable limits of the model and,  on the other hand,  the low energy regime can be approximated by a quadratic integrable Hamiltonian, as small oscillations around the minimal energy configuration \cite{Emary03,Bas14B}. The quadratic approximation is  possible for any coupling  except in a vicinity of the critical one, where the quadratic terms vanish in a small oscillation approximation. The latter fact makes possible that, in couplings close to the critical one,  the chaotic region approaches the low energy region. This is exactly what is observed in the cases $\omega= \omega_0$, and $\omega=4\omega_0$, but in the case $\omega=\omega_0/4$, the chaos appears well above the minimal energy for any coupling including the critical one.  

The case $\omega=4\omega_0$ shows an interesting behavior, absent in the other two cases. For $\omega=\omega_0/4$ and $\omega=\omega_0$, chaotic regions appear at large enough excitation energy, even in the normal ($\gamma<\gamma_c$) phase, but in the case $\omega=4\omega_0$, chaotic regions are completely absent in any energy for the normal phase, and chaos appears in large enough  energy as soon as the coupling attains the critical value. Therefore, the breaking of the quadratic approximation at the critical coupling is exhibited as a necessary but not sufficient condition for  the presence of chaos. 

The case $\omega=\omega_0/4$, shown in Fig.\ref{im:ZONAS_L_C}, exemplifies clearly this. It is interesting that in the other two cases, the breaking of the quadratic approximation in the critical coupling,  leads rapidily, as energy is increased, to a chaotic regime, which  seems to point out in the direction of some connection between both phenomena: chaos and QPT. However, the case  $\omega=\omega_0/4$  is a counterexample that indicates that the relation between both phenomena is not as deep as previously thought \cite{Emary03}.

\begin{figure*}
\centering
\begin{tabular}{c c}
 \includegraphics[width= 0.49 \textwidth]{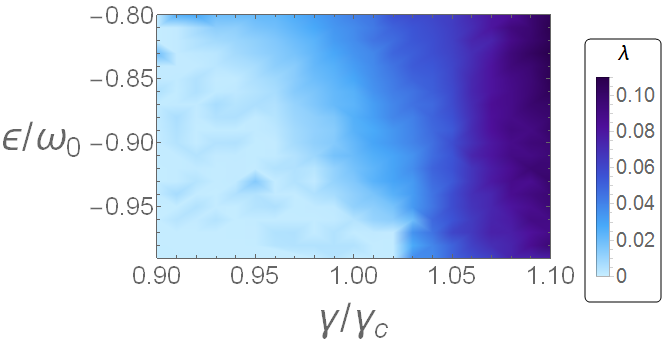} &
\includegraphics[width= 0.49 \textwidth]{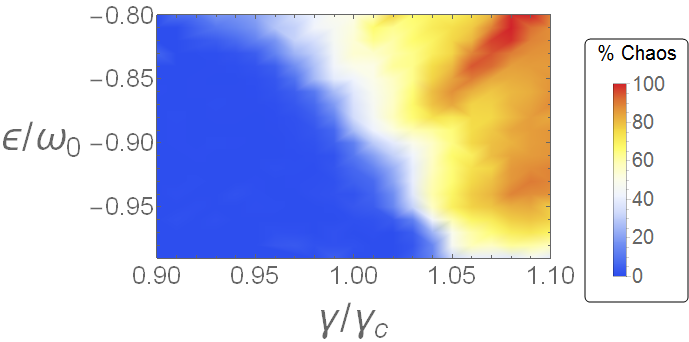}  \\ 
(a) & (b) \\
\end{tabular} 
\caption{\small Zoom around $\gamma=\gamma_c,\epsilon=\epsilon_0$) displaying the Lyapunov exponent over the energy surface (left), and the percentage of chaoticity (right) for  $\omega=\omega_0$.}
\label{fig:por_promL_zoom}
\end{figure*}
In order to go deeper in the analysis of the relation between  chaos and QPT in the Dicke model, and its dependence on the frequencies of the non-interacting modes ($\omega$ and $\omega_0$), in Fig. \ref{fig:por_promL_c} we present the average Lyapunov exponent  (left) and the percentage of chaoticity (right), as functions of the energy  ratio   $\omega/\omega_0$, for $\gamma=\gamma_c$ and different values of the excitation energy $\epsilon$ close to the ground state energy $\epsilon_0$. The ground state energy exactly at the QPT is always $\epsilon_0 = - \omega_0$. The different sets of dots show how much chaos is present at various energies, in the vicinity of  $\omega = \omega_0$. The upper (red) curve shows that for an energy $\epsilon = - 0.8 \omega_0$, there is chaos in a range $\omega \in  (0.8 \omega_0, 1.8 \omega_0)$. Going downward, the following curve (orange), energy $\epsilon = - 0.9 \omega_0$, shows that  the chaotic region has no more than 20\% of chaotic trajectories, and they are restricted to a range $\omega \in  (0.9 \omega_0, 1.4 \omega_0)$. For energies lower than $\epsilon = - 0.95 \omega_0$ the trajectories are all regular, there is no chaoticity.

Another way of studying the presence of chaos in a close neighborhood of the QPT is to fix the the frequencies of the non-interacting modes  in resonance ($\omega = \omega_0$), and perform a detailed study of the region of energies close to the ground state, and coupling constant $\gamma$ close to its critical value $\gamma_c$. These are zooms of the previous maps, and are shown in Fig. \ref{fig:por_promL_zoom}. It confirms that the region around the QPT is {\bf regular}, and that there is no direct relationship between the QPT and the onset of chaos, even in resonance. 

The dependence of the onset of chaos on the frequencies of the non-interacting modes was expected, because, as discussed above, the limit cases $\omega_o\rightarrow 0$ or $\omega\rightarrow 0$ yield integrable Hamiltonians, but what it was not expected is that the presence of chaos at low excitation energy at the critical coupling disappears rapidly as soon as the resonance condition is not fulfilled.

\subsection{Chaos and ESQPT}
As mentioned in the introduction, in \cite{Per11} it is suggested that the onset of chaos in the Dicke model is more related  with the ESQPT than with the QPT. The two  ESQPTs at energies $\epsilon=\pm \omega_0$ are indicated in the maps of Fig. \ref{im:ZONAS_L_C} by horizontal dashed lines. As in the case of the QPT the relation between the onset of chaos and ESQPTs is rather weak, even if in some particular  cases ($\omega=\omega_0$ and $\omega=4\omega_0$) a close  relation seems to be present. It is in the  resonant case $\omega=\omega_0$ where a more direct relation  seems to appear. In this case, even if chaotic regions appear well below the energy of the ESQPT ($\epsilon<\epsilon_c=-\omega_0$) for $\gamma>\gamma_c$, the onset of hard chaos (100 \% of chaos) occurs at energies close to the critical one ($\epsilon_c=-\omega_0$). However, in the other two cases, particularly in the case $\omega=\omega_0/4$ the onset of chaos, even the hard chaos regime, seems to have nothing to do with the ESQPT. The main conclusion is that the relation between the  onset of chaos (and hard chaos) in the model and the ESQPT   is strongly dependent on the parameters of the model and no general deeper relation can be established.

\section{Conclusions}

A global survey of the dynamics of the classical Dicke model Hamiltonian was performed.  The classical Hamiltonian is the one that  results   from the semiclassical approximation in terms of coherent states. We have focused on the onset of chaos in the model as a function of energy and coupling for three sets of frequencies of the atomic (qubit) and field (bosonic) modes, the resonant and two off-resonant cases. The  percentage of chaos and mean Lyapunov exponent over the available phase space were calculated for  a wide range of energies and atom-field couplings, which allowed us to explore the different dynamical regimes, from regular to fully chaotic. The relation between the onset of chaos and critical phenomena (QPT and ESQPT) of the model was discussed. One of the main conclusions is based on the study of the case $\omega=\omega_0/4$, where the appearance of chaos takes places in regions far away from those where the QPT and the ESQPT occur.  
It exhibits that the simultaneous occurrence of both chaos and quantum phase transitions depends strongly on the Hamiltonian parameters, and are not intrinsically related.
 In the literature, the more studied case is the resonant one ($\omega=\omega_0$), where the numerical results  seemed to indicate the existence of a deep relation between chaos, QPT and ESQPT,  but a close look at the vicinity of the QPT in resonance shows that there is always a regular region around the QPT and that the presence of  chaos at low excitation energies at the critical coupling, occurs only for a small interval around the resonant $\omega=\omega_0$ case .

Thanks to the outstanding classical and quantum correspondence in terms of chaos and regularity, the complete maps of chaos presented are a very  convenient guide to future studies where the kind of dynamics of the classical model is useful to know in advance the behavior of the respective quantum version of the model. Such studies include quench dynamics \cite{Per111}, equilibration and thermalization of the Dicke model \cite{Alt121,Alt122,IGM15}.     
    
\section*{Acknowledgements}
J.Ch.C, M.A.B.M. and J.G.H. thank the hospitality and the interesting conversations with P. Cejnar and P. Stransky in Prague. M.A.B.M. thanks D. Wisniacki his valuable suggestions. This work has received partial economical support from Consejo Nacional de Ciencia y Tecnolog\'ia (Conacyt): SEP-Conacyt  and RedTC-Conacyt, Mexico. 

\appendix 
\section{Lyapunov Characteristic Exponent (LCE)} \label{Ap:LCE}

The LCE are asymptotic measures characterizing the average rate of growth (or shrinking) of small perturbations along the solutions of a dynamical system. The concept was introduced by Lyapunov when studying the stability of non-stationary solutions of ordinary differential equations \cite{Lyapunov92,Oseledets68,Benettin80,Parker89,Cencini10,Skokos10,Dubeibe13} and has been widely employed in dynamical systems since then.
\subsection{Variational equations} \label{Ap:Eq_var}
In regular, non-chaotic  systems, the distance $\Vert\delta\textbf{x}(t)\Vert$ between a given trajectory and another one, built from a small perturbation in the initial conditions, remains close to zero, or increase at most algebraically as time evolves. In chaotic systems this distance diverges exponentially in time.
\begin{equation}
\Vert\delta\textbf{x}(t)\Vert \sim  e^{\lambda t} \Vert\delta\textbf{x}(0)\Vert
\end{equation}
The parameter characterizing this instability along the path can be defined by taking the double limit
\begin{equation}
\lambda=\lim_{t\rightarrow\infty}\lim_{\Vert \delta\textbf{x}_{0}\Vert\rightarrow 0}\;\dfrac{1}{t}\ln\dfrac{\Vert\delta\textbf{x}(t)\Vert}{\Vert \delta\textbf{x}_{0}\Vert}
\label{ec:LCE}
\end{equation}
When this limit exists and is positive, the trajectory is extremely sensible to the initial condition and is called chaotic. The parameter $\lambda$ is the \textit{Lyapunov Characteristic Exponent} (LCE).\\
To obtain the LCE in autonomous system it is necessary to solve the dynamical equations $F(\textbf{x})$ and the \textit{fundamental matrix} $\boldsymbol{\Phi}_{t}$ simultaneously (for more details see \cite{Parker89,Skokos10}),
\begin{equation}
\begin{pmatrix}
\dot{\textbf{x}} \\
\dot{\boldsymbol{\Phi}} \\
\end{pmatrix}=\begin{pmatrix}
F(\textbf{x}) \\
D_{\textbf{x}}F(\textbf{x})\boldsymbol{\Phi}\\\end{pmatrix},
\label{FdF1}
\end{equation}
where $D_{\textbf{x}}F(\textbf{x})$ is the \textit{Jacobian matrix}, 
with the initial conditions
\begin{equation}
\begin{pmatrix}
\textbf{x}(t_{0}) \\
\boldsymbol{\Phi}(t_{0}) \\
\end{pmatrix}=\begin{pmatrix}
\textbf{x}_{0} \\
\mathbb{I}\\\end{pmatrix}.
\label{ec:IC}
\end{equation}

In this form the perturbation $ \delta\textbf{x}_{0}$ of $\textbf{x}_{0}$ is given as
\begin{equation}
\delta\textbf{x}(t)=\boldsymbol{\Phi}_{t}(\textbf{x}_{0})\cdot\delta\textbf{x}_{0}.
\end{equation}

\subsection{Stability and critical points in Dynamics of Classical Dicke Model }

In the Dicke model, $\textbf{x}=(p,q,P,Q)$, with these four generalized coordinates defined in the main text.
Using equation (\ref{ec:haclPQ}), the dynamical equations $F(\textbf{x})$ and the Jacobian matrix $D_{\textbf{x}}F(\textbf{x})$ are 
\begin{widetext} 
 \begin{equation}
F(\textbf{x})=\left\{
\begin{array}{c}
-\gamma  Q\, \sqrt{4-P^2-Q^2}-q\, \omega \\
p\, \omega \\
\frac{\gamma \, q
   Q^2}{\sqrt{4-P^2-Q^2}}-\gamma \, q\, \sqrt{4-P^2-Q^2}-Q\,\omega_0\\
   P\,\omega_0-\frac{\gamma  P q Q}{\sqrt{4-P^2-Q^2}} \\
\end{array},
\right.
\label{eq:Din_PQ}
\end{equation}
\begin{equation}
D_{\textbf{x}}F(\textbf{x})=\left(
\begin{array}{cccc}
 0 & -\omega  & \frac{\gamma  P Q}{\sqrt{-P^2-Q^2+4}} & \frac{\gamma  \left(P^2+2
   Q^2-4\right)}{\sqrt{-P^2-Q^2+4}} \\
 \omega  & 0 & 0 & 0 \\
 0 & \frac{\gamma  \left(P^2+2 Q^2-4\right)}{\sqrt{-P^2-Q^2+4}} & -\frac{\gamma  P
   \left(P^2-4\right) q}{\left(-P^2-Q^2+4\right)^{3/2}} & -\frac{\gamma  q Q \left(3 P^2+2
   \left(Q^2-6\right)\right)}{\left(-P^2-Q^2+4\right)^{3/2}}-\omega_0 \\
 0 & -\frac{\gamma  P Q}{\sqrt{-P^2-Q^2+4}} & \frac{\gamma  q Q
   \left(Q^2-4\right)}{\left(-P^2-Q^2+4\right)^{3/2}}+\omega_0 & \frac{\gamma  P
   \left(P^2-4\right) q}{\left(-P^2-Q^2+4\right)^{3/2}} \\
\end{array}
\right).
\label{eq:J_PQ}
\end{equation}
\end{widetext}
The critical points of the dynamical system correspond to $F(\textbf{x})=\bar{0}$. Employing Eq.(\ref{eq:Din_PQ}),  four critical points: $\textbf{x}_{c0}=(0,0,0,0)$, $\textbf{x}_{c\pi}=(0,0,-2\sin\phi,2\cos\phi)$ and  $\textbf{x}_{\pm c}=\left(0,\pm\dfrac{2\gamma^2}{\omega}\sqrt{1-\dfrac{\gamma^4_c}{\gamma^4}},\; 0,\;\mp\sqrt{2\left( 1-\dfrac{\gamma^2_c}{\gamma^2}\right) }\right) $.
The last two critical points can only exist in the super radiant region 
$\gamma\geq\gamma_c$.
Their energies are $h_{cl}(\textbf{x}_{c0})=-\omega_0$, $h_{cl}(\textbf{x}_{c\pi})=\omega_0$ and  $h_{cl}(\textbf{x}_{\pm c})=\epsilon_0(\gamma)$, defined in Eq. (\ref{eq:gse}).

\section{Regular and chaotic trajectories} \label{Ap:LCE_Sample}

In this Appendix we present a few representative examples of regular and chaotic trajectories and their associated Poincar\'e sections. The case selected has $\epsilon=-1.4\,\omega_0$ and $\omega=\omega_0$, with a mixed phase space. Two initial conditions are shown, one regular, with $\lambda=\;0$, and the second one chaotic, with $\lambda\sim\;0.05$.

\subsection{Lyapunov exponent $\lambda=\;0$}

\begin{figure}
 \centering
  \begin{tabular}{c c}
    \includegraphics[width= 0.2 \textwidth]{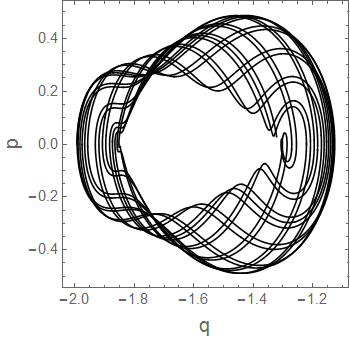} & \includegraphics[width= 0.2 \textwidth]{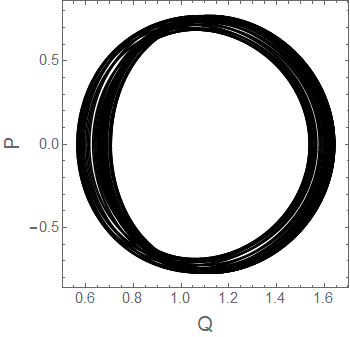}\\ 
    (a) & (b)\\
  \end{tabular}
   \caption{\small Trajectories with initial condition $\textbf{x}_r$ in the canonical variables ($q,p$) and ($Q,P$).}
   \label{im:tray_l_0}               
\end{figure}

The trajectory with initial conditions $\textbf{x}_r=(0,q(\epsilon),0,0.707)$ 
exhibits regular dynamics, as can be seen in the projection of the variables ($q,p$) and ($Q,P$) shown in Figure \ref{im:tray_l_0}.\\ 
\begin{figure}
   \includegraphics[width= 0.2 \textwidth]{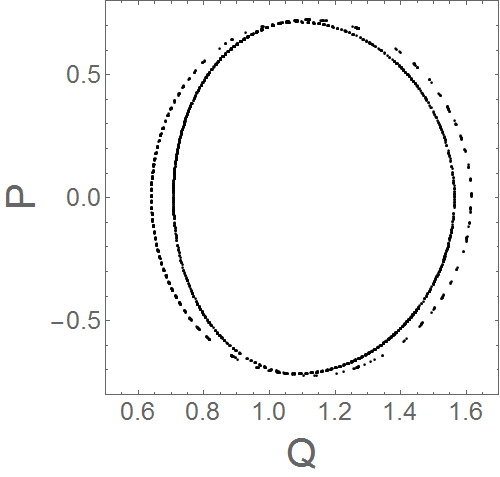} 
   \caption{\small Poincar\'e section for the trajectory with initial condition $\textbf{x}_r$ in the canonical variables ($Q,P$).}
   \label{im:Poinc_l_0}               
\end{figure}
The corresponding Poincar\'e surface section for the canonical variables ($Q,P$) is presented in Fig. \ref{im:Poinc_l_0}. It is restricted to an annular area in phase space, qualitatively identifiable as regular.
The maximum Lyapunov exponent along this trajectory is 0.0007, smaller than the numerical limit $\lambda_{min}= 0.002$, and for this reason associated with a null Lyapunov exponent, identifying it as a regular orbit.

\subsection{Lyapunov exponent $\lambda\sim\;$0.05}

\begin{figure}
  \centering
  \begin{tabular}{c c}
    \includegraphics[width= 0.2 \textwidth]{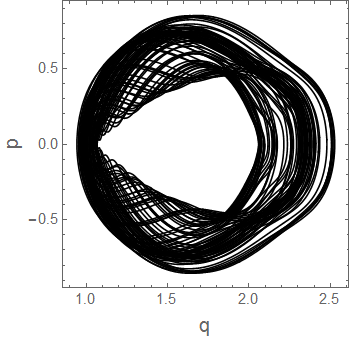} & \includegraphics[width= 0.2 \textwidth]{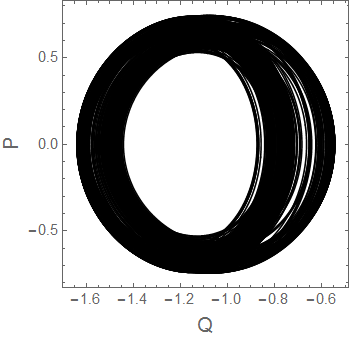}\\ 
    (a) & (b)\\
  \end{tabular}
   \caption{\small Trajectories with initial condition $x_{c}$ in the canonical variables ($q,p$) and ($Q,P$).}
   \label{im:tray_l_05}               
\end{figure}

The trajectory with initial conditions $\textbf{x}_c=(0,q(\epsilon),0.6481,-1.371)$ has Lyapunov exponent $\lambda\sim\;0.05$. 
Its tendency to fully cover the available phase space can be inferred from the projections of the trajectory on the $(q,p)$ and $(Q,P)$ planes shown in Fig. \ref{im:tray_l_05}.
The Poincar\'e section corresponding to this trajectory 
in the space $(Q,P)$ is presented in Figure \ref{im:tray_l_05}.
\begin{figure}
   \includegraphics[width= 0.2 \textwidth]{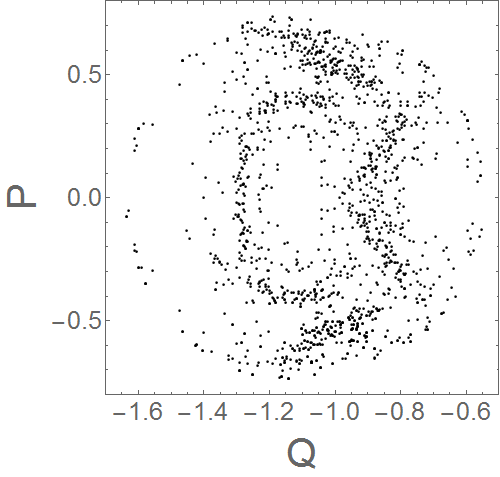} 
   \caption{\small Poincar\'e section for the trajectory with initial condition $\textbf{x}_c$ in the canonical variables ($Q,P$).}
   \label{im:Poinc_l_05}               
\end{figure}
The presence of scattered point covering the area qualitatively characterizes the chaotic behavior. Its Lyapunov exponent is $\lambda\sim\,$ 0.05, definitely larger than the cut 0.002.

\section{Evaluation of the Lyapunov exponents}

\subsection{Separation between trajectories}

In the Dicke model the phase space for the four generalized coordinates $(p,q,P,Q)$ is bounded.  The \textit{geometric distance} between two trajectories $\textbf{x}_1(t)$ and  $\textbf{x}_{2}(t)$, $dx(t) \equiv \mid \textbf{x}_2(t) -\textbf{x}_{1}(t)\mid$,  can grow in time up to a maximum value, and after that can only oscillate around it. On the other hand, the \textit{separation in the tangent space}  $\delta \textbf{x}(t)$, defined in Eq. (B5), is not bounded, and is the one employed in the evaluation of the Lyapunov exponents. 

\begin{figure}
   \centering
  \begin{tabular}{c c}
    \includegraphics[width= 0.23 \textwidth]{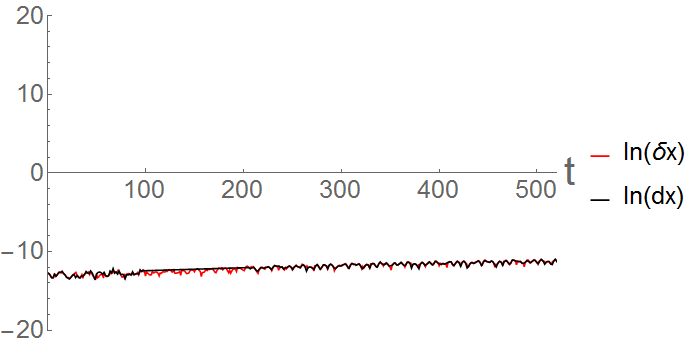} & \includegraphics[width= 0.23 \textwidth]{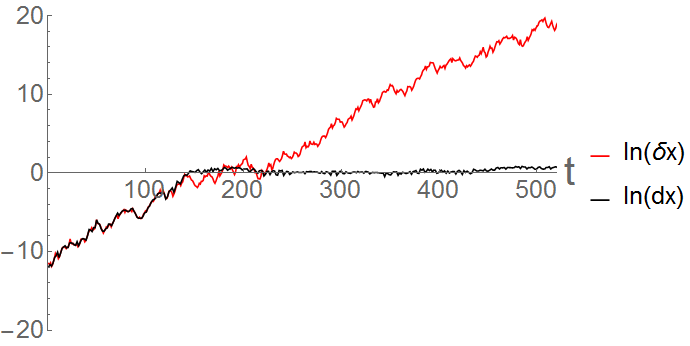}\\ 
    regular & chaotic\\
  \end{tabular}
  \caption{\small  (Color online) Geometric distance $dx(t)$ (black) and separation $\delta x(t)$ (red) in logarithmic scales between two close trajectories, in a regular regime  (left) and  a  chaotic one (right).}
   \label{im:sep_l}                
\end{figure}

In Fig. \ref{im:sep_l}  these two separations are shown, in logarithmic scale, for a regular (left) and a chaotic (right) trajectory.  The separation in the initial conditions is $dx_{0}\sim 10^{-6}$.
In the regular case the two trajectories remain quite close to one another, and both distances are small, and hard to distinguish, at any time. The chaotic trajectories clearly diverge. Their geometric distance saturates at  $t \approx 150$, while the separation in the tangent space keeps growing exponentially. 

\begin{figure}
  \centering
  \begin{tabular}{c c}
    \includegraphics[width= 0.23 \textwidth]{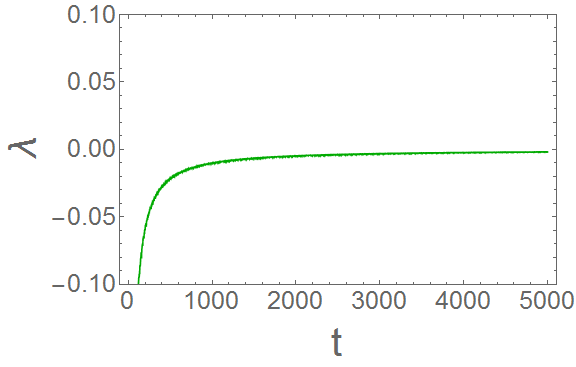} &     \includegraphics[width= 0.23 \textwidth]{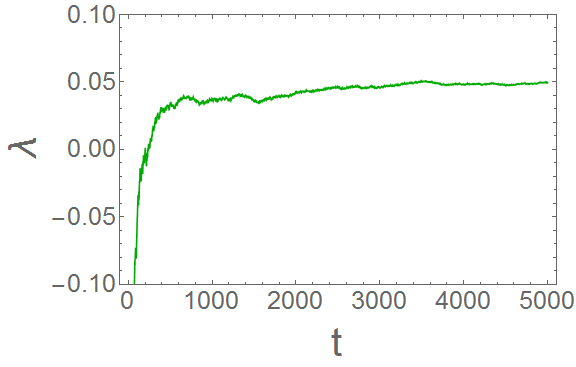} \\
     regular & chaotic\\
  \end{tabular}
   \caption{\small  (Color online)  Convergence of the Lyapunov exponents as functions of time, for regular (left) and chaotic (right) trajectories.}
   \label{im:conv_sep}                
\end{figure}

The slope of the separation in the tangent space, in logarithmic scale, defines the Lyapunov exponents. Their numerical values, estimated with a linear regression in the interval $(0,t)$, are displayed in Fig.  \ref{im:conv_sep} for times up to 5000, for a regular (left) and a chaotic (right) trajectory. The Lyapunov exponent of the regular trajectory converges to $\lambda =0.0007$ (consistent with a zero Lyapunov exponent), and for the chaotic trajectory to $\lambda = 0.05201$. For both trajectories, times of order 500 are enough to obtain the converged values.            

\subsection{Sampling the vicinity of a given initial condition}\label{sampling}

\begin{figure}
 \centering
\begin{tabular}{c c}
\includegraphics[width= 0.22 \textwidth]{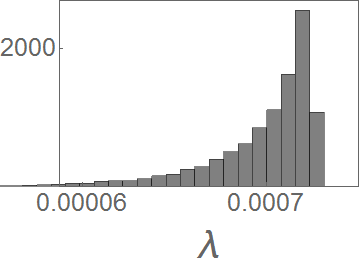} &
\includegraphics[width= 0.24 \textwidth]{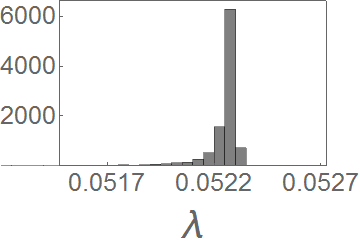} \\ 
     regular & chaotic\\
\end{tabular}
 \caption{\small Histograms of values obtained in the sampling of Lyapunov exponents, for regular (left) and chaotic (right) trajectories.}   
 \label{im:muest_lrc}                
\end{figure}

In the above subsection an example was given of two trajectories whose distances  diverge exponentially with time, and  other two which remain close. To estimate the Lyapunov exponent for a given initial conditions, thousand of trajectories are selected, whose initial condition is chosen in a small vicinity of the one under  study. In Fig.  \ref{im:muest_lrc} the histograms of the distributions of the Lyapunov exponents are displayed. They were obtained using 10,000 randomly selected initial conditions in a neighborhood with $dx\sim 10^{-6}$ around the initial condition, for the same two trajectories of the above subsection, one regular and one chaotic.   The stability of the method is clearly confirmed, as the distributions are confined to a small region in the value of the Lyapunov exponent, with dispersion in the fourth decimal digit. \\

\subsection{Lyapunov exponents along a trajectory}

\label{Ap:LCE in time}
 \begin{figure*}
 \centering
  \begin{tabular}{c c c}
   \includegraphics[width= 0.3 \textwidth]{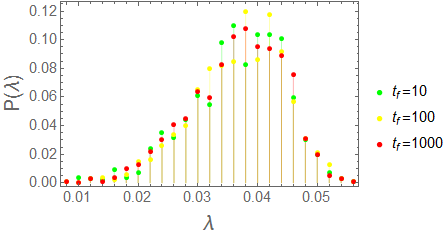} & \includegraphics[width= 0.3 \textwidth]{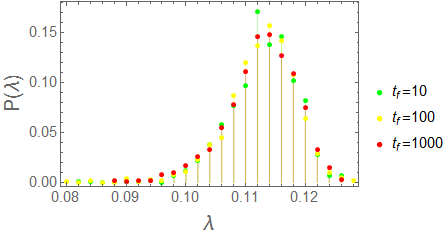} & \includegraphics[width= 0.3 \textwidth]{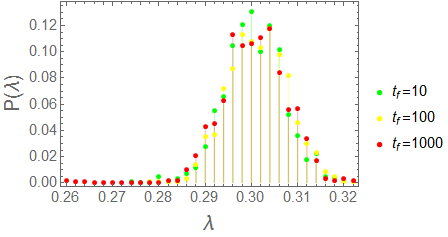} \\ 
(a) & (b) & (c)\\  
  \end{tabular}
  \caption{\small  (Color online) Probability distribution of the Lyapunov exponents, evaluated for three different energies  $\epsilon=-1.4\,\omega_0$ (a), $\epsilon=-1.1\,\omega_0$ (b) and $\epsilon=-0.5\,\omega_0$ (c), for three time scales $t_{f}$  for each trajectory, with initial conditions $\bar{x}(0)=(0, q(\epsilon), 0, 0.948)$.}
   \label{im:Lyap_Trayec}                
\end{figure*} 

As the Lyapunov exponents depend on the initial conditions, they can be evaluated for different points along a given trajectory. The values obtained are in general close, with a bell-shaped distribution, as shown in Fig. \ref{im:Lyap_Trayec}. The three examples presented are chaotic, with energies  $\epsilon=-1.4\,\omega_0$ (a), $\epsilon=-1.1\,\omega_0$ (b) and $\epsilon=-0.5\,\omega_0$ (c), and Lyapunov exponents $\lambda \approx 0.04 \pm 0.01, 0.11 \pm 0.01$ and $0.30 \pm 0.01$, respectively. The points along the trajectories are evaluated in three time scales, from $t_i=0$ to $t_f=10, 100$ or $1000$, displayed with different colors.  It can be seen that the distributions are insensitive to the time scale, and their dispersion width is close to $0.01$.


\end{document}